\documentclass{aa}
\usepackage{txfonts}
\usepackage{graphicx}
\usepackage{amsmath}
\usepackage{natbib}
\usepackage{amssymb}
\bibpunct{(}{)}{;}{a}{}{,} 
\begin{document}
\title{Scale invariant cosmology III: dynamical  models and comparisons with observations }
\titlerunning{Models and observations}

\author{Andr\'e Maeder
}
\authorrunning{Maeder}

\institute{Geneva Observatory, Geneva University, CH--1290 Sauverny, Switzerland\\
email: andre.maeder@unige.ch
}

\date{Received  / Accepted }


\abstract
{}
{We want to examine the properties of the scale invariant cosmological models, also making the specific  hypothesis 
of the scale invariance of the empty space at large scales.}
{Numerical integrations of the cosmological equations for different values of the curvature parameter $k$ and
 of the  density parameter $\Omega_{\mathrm{m}}$ are performed. We compare the  dynamical properties of the models to the observations at different epochs.   }  
{The main numerical data and graphical representations  are given for models computed with different
curvatures and density-parameters.
 The models with non-zero density start explosively with first a braking phase followed by a continuously  accelerating expansion.
The comparison of the models with the recent observations from supernovae SN Ia, BAO and CMB data from Planck 2015
shows that the scale invariant model with $k=0$ and $\Omega_{\mathrm{m}}=0.30$  very well fits  the observations in the usual
$\Omega_{\mathrm{m}}$ vs. $\Omega_{\Lambda}$ plane and 
consistently accounts for the accelerating expansion or dark energy. 

 The expansion history is   compared to observations 
in the plot $H(z)$ vs. redshift $z$, the  parameters $q_0$ is also examined, as well the recent data about
 the redshift $z_{\mathrm{trans}}$
of the transition between braking and acceleration. These dynamical tests are fully satisfied by the scale invariant models.
The past evolution of matter and radiation density is  studied, it shows   small differences with respect to the standard case. }
{These first comparisons   are  encouraging further investigations on scale invariant cosmology with the
assumption of scale invariance of the empty space at large scales. }


\maketitle
 
 \section{Introduction}
 
 In the two previous papers of this series, we have derived   the equations of a scale invariant cosmology and studied their properties. 
 Two  tentative, but fundamental, hypotheses are at the basis of these works. The first is that we may  apply a general equation of the
 gravitational field, which in addition to the general covariance of General Relativity  also possesses the property of scale invariance.
 Developments along this line were already performed in the past by \citet{Eddi23,Dirac73,Canu77}. The second hypothesis is that the empty space, for exemple 
 in the sense it is used in the  Minkowski metric, should be scale invariant at macroscopic and large scales. It means that if, at such scales,
 we extend or contract  the empty space, its properties are  still the same.  
  This hypothesis, which as far we know is new in this context, allows us to establish some differential  equations
  connecting the scale factor $\lambda(x^{\mu})$  and the Einstein cosmological constant
 $\Lambda_{\mathrm{E}}$, this leads to relation (\ref{lamb}).  It also brings constraints on the scale factor and  useful simplifications in the scale 
 invariant equations.

 The two above hypotheses lead to  far-reaching consequences in physics and cosmology. The basic equations of cosmology are modified, showing 
 an acceleration of the expansion after a certain initial period, the duration of which depends on the mean density of the Universe. Another
 major consequence of  the scale invariance is that the laws of conservation of matter-energy  show some dependence on the cosmic
 time. This dependence is very weak for models with a non-zero matter density, but at the conceptual level this is not a minor effect. 
 
 We do think it is worth to undertake the present exploration for two main reasons. One is that  the recent cosmological results
 suggest that a totally unknown form of matter-energy, the dark energy, dominates the energy content of the Universe. This is
 a major problem.
The other main  reason is that   scale invariance is not a kind of adjusted trick to make  things work. But it is a basic physical change,
that responds to the  fundamental wish \citep{Dirac73} that the equations expressing basic laws should be invariant under the widest group
of transformations. 

In this work, we construct the corresponding cosmological models, examine their dynamical properties and make close comparisons with observations.
If there is no disagreement, this may be considered as  encouraging, studies and comparisons will have to be pursued. 
If we find some serious disagreement, we may turn to the conclusion  
that at least one, or maybe the two fundamental hypotheses we have made do not correspond to the reality of Nature.

In Section 2, we express  the equations of cosmology in an integrable form. In Section 3, 
we find and discuss the numerical solutions of the scale invariant  models for the flat case with $k=0$, while
 the cases with the  curvature parameter $k=\pm 1$ are analyzed in Section 4.  Section 5 is devoted to the comparisons of
 models and observations, in particular the density parameters and the Hubble constant at present. In Section 6, we perform some 
 dynamical tests at other epochs concerning the Hubble parameter $H(z)$ vs. $z$, 
 the value of the deceleration parameter $q_0$ and the transition from braking to acceleration.
 In Section 7,  the evolution of the matter and radiation densities, as well as the temperature over the ages are derived.
 Section 8 contains the conclusions.

 \section{Scale invariant cosmological models}  \label{models}

Scale invariance is the invariance to a transformation of the line element like
$ ds' \, = \, \lambda(x^{\mu}) \, ds $,
where $ds'\,^2  =  g'_{\mu \nu}\, dx^{\mu} \, dx^{\nu} $ is the line element in the framework of General Relativity, 
while  $ds^2  =  g_{\mu \nu}\, dx^{\mu} \, dx^{\nu} $  
is the line element in a  more general framework where scale invariance is  a  property. 
  The scale factor $\lambda$ only depends on the cosmic time in agreement with the Cosmological Principle.
 The scale invariance of the empty space at large scales
 implies  a solution  for $\lambda$ of the form 
\begin{equation}
\lambda \, = \, \sqrt{\frac{3}{\Lambda_{\mathrm{E}}}} \, \frac{1}{c \,t}  \, .
\label{lamb}
\end{equation}

\noindent
If we  take $\lambda$ to be unity at the present cosmic time   $t_0$, we get $\lambda \, = \, t_0/t$
and thus $ \dot{\lambda}/  \lambda= -1/t$.
    The constraint on the choice of the origin of $t$ will come from the chosen cosmological models. Origins at time $t_{\mathrm{in}}$  larger than 0  considerably  reduce
the amplitude of the variations of the scale factor $\lambda$ over the ages.

The  basic  equations of  the scale invariant cosmology, that we derived from the above two fundamental hypotheses are 
according to Paper II,

\begin{equation}
\frac{8 \, \pi G \varrho }{3} = \frac{k}{R^2}+\frac{\dot{R}^2}{R^2}+ 2 \,\frac{\dot{R} \dot{\lambda}}{R \lambda} \, \; ,
\label{E1}
\end{equation} 
\noindent
and
\begin{equation}
-8 \, \pi G p  = \frac{k}{R^2}+ 2 \frac{\ddot{R}}{R}+\frac{\dot{R^2}}{R^2}
+ 4 \frac{\dot{R} \dot{\lambda}}{R \lambda}  \, \; .
\label{E2}
\end{equation}
\noindent
The combination of these two equations leads to

\begin{equation}
-\frac{4 \, \pi G}{3} \, (3p +\varrho)  =  \frac{\ddot{R}}{R} + \frac{\dot{R} \dot{\lambda}}{R \lambda}  \,   \,.
\label{E3}
\end{equation}

\noindent
 The gravitational  constant $G$ is  a true constant,  $k$ is the curvature parameter
  ($0$ and $\pm 1$),  $p$ and $\varrho$ are the pressure and density in the scale invariant system. 
  In these equations, we have also explicitly  accounted  for the scale invariance of the empty space at large scales.
  Compared to the standard equations of Friedman models, the above ones only differ by the presence of a term in 
 $ \dot{R} \, \dot{\lambda}/ (R \, \lambda)$, which represents  an acceleration opposed to gravitation,
 since $ \dot{\lambda}/  \lambda$ is negative, as seen above.

   The solutions of these equations depend  on the equation of state of the medium we are considering.
   For an equation of state of the form

\begin{equation}
P \, = \, w \,  \varrho \, ,  \quad  ( \mathrm{with \;} c^2 =1) \,  ,
\label{etat}
\end{equation}
\noindent
where $w$ is a constant,
the first two equations lead to a first integral

\begin{equation}
\varrho \, R^{3(w+1)}  \,  \lambda ^{(3w+1)} \,= const. 
\label{3w}
\end{equation}
\noindent
as shown in Paper II.  For the case $w=0$ of ordinary matter of density
$\varrho_{\mathrm{m}}$, exerting no pressure, we get   $\varrho_{\mathrm{m}} \, \lambda \, R^3 =const.$  
 If $\lambda(t)$ is a constant, one gets the usual equations of cosmologies for the expansion term $R(t)$.  In Section 7, we also consider the
 phase of the Universe evolution where radiation is dominating.

We are first searching the solution  of  the cosmological equations  for the case of ordinary  matter  
with density  $\varrho_{\mathrm{m}}$
and $w=0$. We start from
(\ref{E1}) and multiply it by $R^3 \lambda$ so that we may use the above first integral of the equation of state, 

\begin{equation}
\frac{8 \, \pi G \varrho_{\mathrm{m}} R^3 \lambda}{3} = k \, R \lambda +\dot{R}^2 R \lambda+ 2 \, \dot{R} R^2 \dot{\lambda} \,
\label{E11}
\end{equation} 
\noindent
The first member is a constant. With  $\lambda=t_0/t$ and choosing the timescale such  that  at present $t_0=1$, we have

\begin{equation}
\dot{R}^2 R \, t - 2 \, \dot{R}\, R^2 + k \, R \, t - C \, t^2 =0 \, , 
\label{E12}
\end{equation}
\noindent
with
\begin{equation}
 C=\frac{8 \, \pi G \varrho_{\mathrm{m}} \, R^3 \lambda}{3} \, .
 \label{C}
 \end{equation}
\noindent
Eq. (\ref{E12}) is a differential equation of order 1 and degree 2. 
The time $t$ is expressed in units of the present time $t_0$ taken equal to 1, at which we also
assume $R_0 = 1$. The origin, the Big-Bang if any one, occurs when $R(t)=0$ at an initial time 
$t_{\mathrm{in}}$ which is not necessarily 0. Indeed, the cosmological models below will show 
that it is only in the case of an empty Universe (cf. Paper II), that
the origin appears to lie  at $t_{\mathrm{in}}=0$. We  notice that if we have a  solution $R$ vs. $t$,
 then $(x \, R)$ vs. $(x \, t)$ is also a  solution, thus the solutions are also scale invariant, 
 as expected from our initial assumptions. 
To integrate this equation, we need to have numerical values  of $C$, corresponding to different values of the
density in the model Universe. The way of treating the problem depends on
the curvature parameter $k$.

\section{Cosmological models with a flat space ($k=0$)}   \label{flat}

\noindent
The case of the Euclidean  space is evidently the most interesting one in view of the confirmed  results of the  space missions
investigating the Cosmic Microwave Background (CMB) radiation with 
Boomerang  \citep{deBern00}, WMAP \citep{Benn03} and the  \citet{Planck15}. Expression (\ref{E12})
becomes

\begin{equation}
\dot{R}^2 R \, t - 2 \, \dot{R}\, R^2  - C \, t^2 =0 \, ,
\label{E13}
\end{equation}
\noindent
In $t_0=1$  and  $R_0=1$, with  the Hubble constant at the present time $H_0 =\dot{R_0}/R_0$, the above relation   leads to

\begin{equation}
H^2_0 -2 \, H_0  = C \, .
\label{ch}
\end{equation}
\noindent
This allows  us to  express $H_0$ as a function of $C$ with 

\begin{equation}
H_0 \, =  \, 1  \pm  \sqrt{1+C} ,
\label{hc}
\end{equation}
\noindent
where we take the sign  + since $H_0$ is always positive.

\begin{figure*}[t!]
\centering
\includegraphics[width=.95\textwidth]{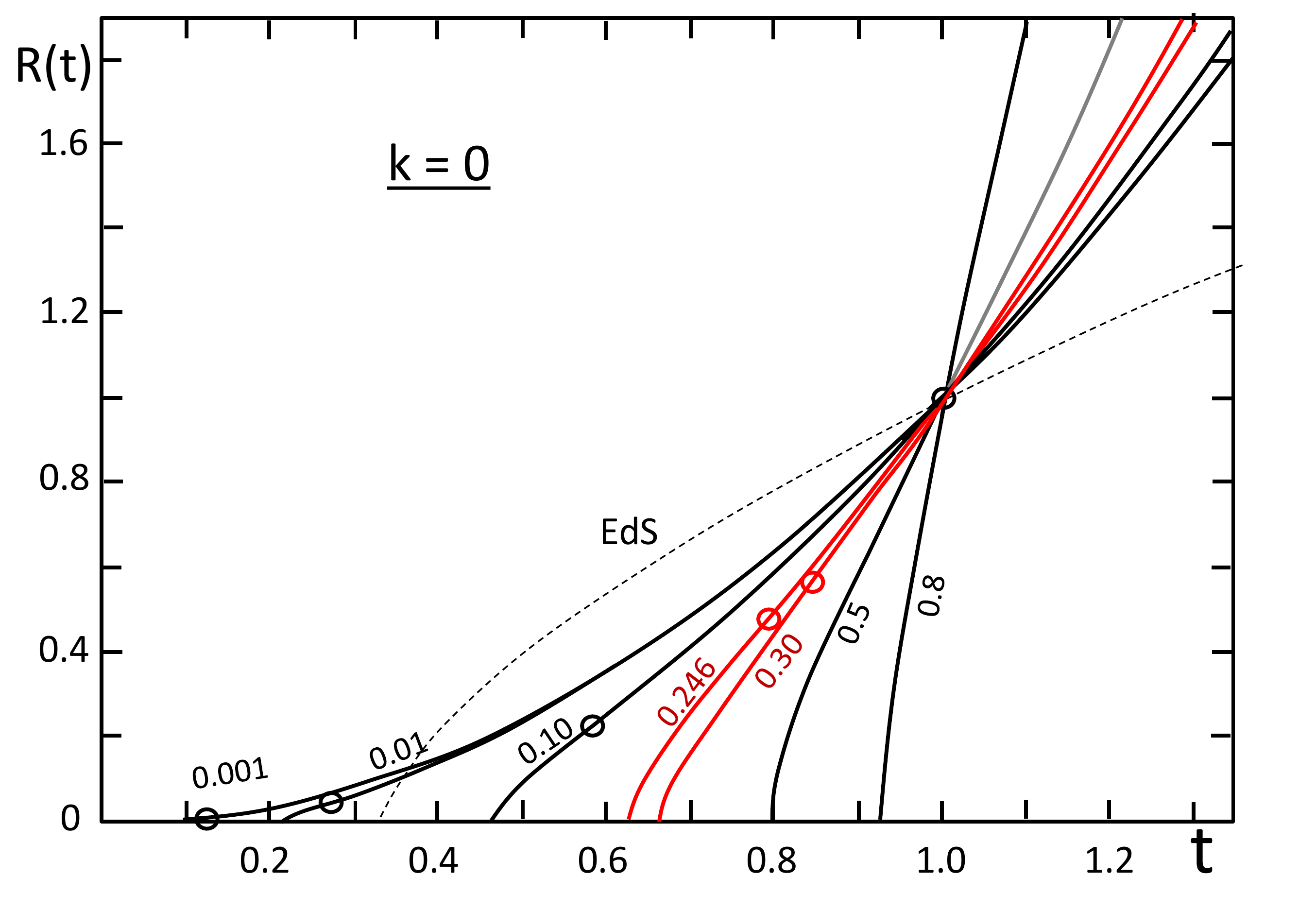}
\caption{Some solutions of R(t) for the models with $k=0$ and $\Omega^*_{\mathrm{m}}=1$. The curves are labeled by the  values of $\Omega_{\mathrm{m}}$, the usual density parameter defined by (\ref{oprime}) and considered at the present time $t_0$. 
 The Einstein-de Sitter model (EdS) is indicated by a dotted line. The small circles on the curves show the transition point between braking
$(q >0) $ and  acceleration $(q<0)$, for  $\Omega_{\mathrm{m}}=0.80$, this point is at $R= 2.52$. The two  red curves indicate  models corresponding to the observational values  of
$\Omega_{\mathrm{m}}=0.246$ \citep{Frie08} and of $\Omega_{\mathrm{m}}=0.30$ given by the \citet{Planck15}.  }
\label{Rtzero}
\end{figure*}

We now want to relate  $C$ to  the density parameters. For $k=0$, the critical matter density  $\varrho^{*}_{\mathrm{ c}}$
at time $t$   is obtained from (\ref{E1}) and the corresponding 
density parameter $\Omega^*_{\mathrm{m}}$ is   defined by 

\begin{equation}
\varrho^{*}_{\mathrm{ c}} =  \frac{3 \, H^2}{8 \pi G}  \left(1- \frac{2}{t \, H }\right)   \quad  \mathrm{and} \; \quad
\Omega^*_{\mathrm{m}} = \frac{\varrho_{\mathrm{m}}}{\varrho^*_{\mathrm{c }}}  \, ,
\label{roc}
\end{equation}

\noindent
as studied in paper II.
 A  remark about the notations:
  we put a * to the  critical density and $\Omega$-parameter defined by (\ref{roc}) to distinguish them from 
  the usual definitions of these parameters, which are

\begin{equation}
\varrho_{\mathrm{c }}= \frac{3 \, H^2}{8 \pi G}  \,  \; \; \quad \mathrm{and} \quad  \Omega_{\mathrm{m}}= \frac{\varrho_{\mathrm{m}}}{\varrho_{\mathrm{c }}}  \, .
\label{oprime}
\end{equation}
\noindent
We  have the following relation between these two $\Omega$-parameters

\begin{equation}
\Omega_{\mathrm{m}} \, = \, \Omega^*_{\mathrm{m}} \left(1- \frac{2}{t \, H }\right)  \, .
\label{deuxo}
\end{equation}
\noindent
The term $\Omega_{\mathrm{m}}$ satisfies at all times the fundamental relation, 

\begin {equation}
\Omega_{\mathrm{m}} \, + \, \Omega_{\mathrm{k}} \, +  \Omega_{\lambda} = \, 1  \, 
 \quad \mathrm{with} \quad  \Omega_{\lambda} = \frac{2}{ H \, t} \, ,
\label{Omegapr}
\end{equation}

\noindent
and  $\Omega_{\mathrm{k}}= -\frac{k}{H^2 \, R^2}$, which is zero here. 
For $k=0$, we have  $\Omega^*_{\mathrm{m}} =1$ at  all times, according to
 (\ref{deuxo}) and  (\ref{Omegapr}).
We have seen in Paper II that except for  of $\Omega^*_{\mathrm{m}} =1$ and 
 $\Omega_{\mathrm{k}}=0$ in the case where $k=0$, the various
 $\Omega$-parameters vary with time in scale invariant models. 
These parameters are generally considered at the present time (as is also the case in the $\Lambda$CDM models).
 This will be the practice generally adopted here, unless explicitly specified.

It is convenient to express  $C$  
(which determines the solution) as a function of   parameter $\Omega_{\mathrm{m}}$.
From  (\ref{Omegapr}), we  have $\Omega_{\mathrm{m}} = 1 - \frac{2}{H_0} $ at the present time $t_0=1$,
 thus  
 
\begin{equation}
H_0 \, = \, \frac{2}{1-\Omega_{\mathrm{m}}} \, .
\end{equation}
\noindent
This expression gives $H_0$ (in unit of $t_0$)  directly from  $\Omega_{\mathrm{m}}$.
We may also now obtain $C$ as a function of $\Omega_{\mathrm{m}}$  at time $t_0$ with the help of (\ref{ch}),

\begin{equation}
C = \frac{4}{(1-\Omega_{\mathrm{m}})^2} -\frac{4}{(1-\Omega_{\mathrm{m}})} =
\frac{4 \, \Omega_{\mathrm{m}}}{(1-\Omega_{\mathrm{m}})^2}  \, ,
\label{copr}
\end{equation}
\noindent
a relation which   allows us to integrate  (\ref{E13})  for a chosen value of the density parameter   $\Omega_{\mathrm{m}}$
at present.

\begin{table*}[t]  
\vspace*{0mm}
 \caption{Cosmological  parameters  of some models with $k=0$ and different $\Omega_{\mathrm{m}} <1 $,
$\Omega_{\mathrm{m}}$ being the usual density parameter at present time $t_0$. 
Note that $\Omega^*_{\mathrm{m}} =1 $ for all these models.
$H_0(t_0)$  is the values of the Hubble constant taking $t_0=1$, $t_{\mathrm{in}}$ is the time when $R(t) = 0$, $ \tau= t_0 -t_{\mathrm{in}}$ is the age of the Universe in units where $t_0=1$, $\tau$(Gyr) is the age of the Universe in Gyr for the considered model assuming that the age of the model with $\Omega_{\mathrm{m}}=0.30$ is 13.8 Gyr,
$H_0(\tau)$ is the Hubble constant in the unit of $\tau$, $t$(q=0) and  $R$(q=0) are the values of $t$ and $R$ at the inflexion point, ``$H_0$ obs`` is the value of the Hubble constant 
in km s$^{-1}$ Mpc$^{-1}$ for an age of the Universe equal to 13.8 Gyr \citep{Frie08}.
} 
\begin{center}  \small
\begin{tabular}{cccccccccccc}
$\Omega_{\mathrm{m}}(t_0)$  &  $C$  & $H_0(t_0)$ & $t_{\mathrm{in}}$ & $q_0$ &  $\tau$ & $\tau$(Gyr)   & $H_0(\tau)$ & 
$t$(q=0) & $R$(q=0) & $\Omega_{\lambda}$ &$H_0$ obs  \\
\hline
 &   &   &   \\
0.001  & 0.0040 & 2.0020 & 0.0999 &- 0.499 & 0.9001 & 37.6   &1.802 & 0.126 & 0.010 & 0.999 & 127.7\\
0.010  & 0.0408 & 2.0202 & 0.2154 & -0.490 & 0.7846 & 32.7   & 1.585 & 0.271 & 0.047 & 0.990 & 112.3 \\
0.100  & 0.4938 & 2.2222 & 0.4641 &- 0.400 & 0.5359 & 22.4   &1.191 & 0.585 & 0.231 & 0.900 & 84.4 \\
0.180  & 1.0708 & 2.4390 & 0.5645 &- 0.320 & 0.4355 &  18.2  &1.062 & 0.711 & 0.364 & 0.820 & 75.3 \\
0.246  & 1.7308 & 2.6525 & 0.6265 &- 0.254 & 0.3735 & 15.6   &0.991 & 0.789 & 0.474 & 0.754 & 70.2 \\
0.300  & 2.4490 & 2.8571 & 0.6694 & -0.200 & 0.3306 &  13.8  &.945 & 0.843 & 0.568 & 0.700 & 67.0\\
0.400  & 4.4444 & 3.3333 & 0.7367 & -0.100 & 0.2633 &  11.0  &0.878 & 0.928 & 0.763 & 0.600 & 62.2\\
0.500  & 8.0000 & 4.0000 & 0.7936 & 0.000 & 0.2064 &    8.6   &0.826 & 1.000 & 1.000 & 0.500 & 58.5 \\
0.800  &  80     &    10    & 0.9282  & 0.300 & 0.0718 &    3.0   &0.718 & 1.170 &  2.520 & 0.200& 50.9 \\
0.990  & 39600  & 200    & 0.9967 &  0.490 & .00335 &    0.14  &0.669  & 1.256   & 21.40  & 0.010& 47.4 \\
\hline
\normalsize
\end{tabular}
\end{center}
\end{table*}

\begin{figure}[h]
\begin{center}
\includegraphics[width=9.6cm, height=7.0cm]{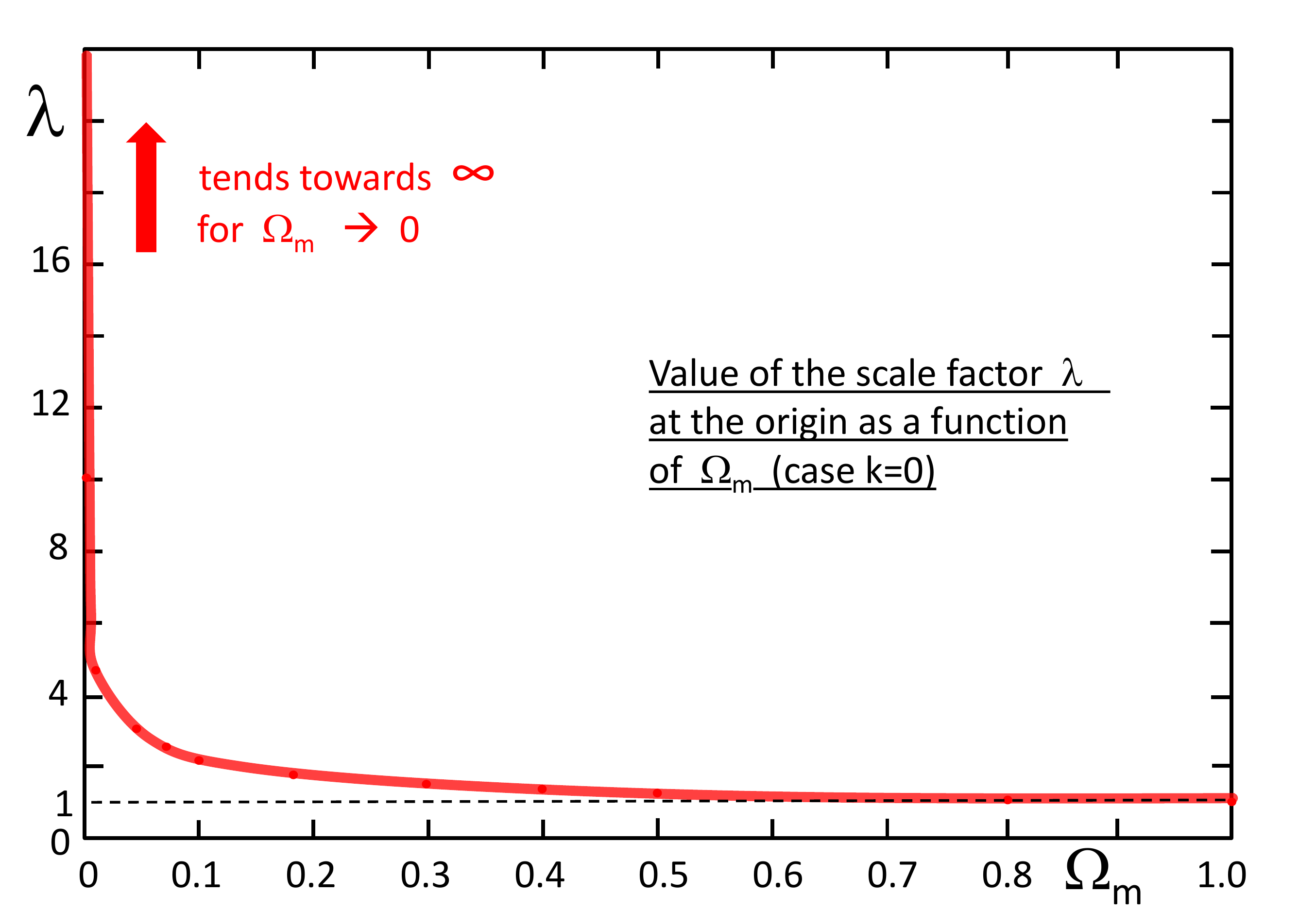}
\caption{Values of the scale factor $\lambda$ at the origin $R(t)=0$ for models with $k=0$ and  
different density parameters $\Omega_{\mathrm{m}}$ at $t_0$. This curve
shows that for increasing densities, the amplitude of the variations of the scale factors $\lambda$ is very much reduced.}
\label{scale}
\end{center}
\end{figure}

 While in the Friedman models, there is a unique value of  the density $\varrho_{\mathrm{c}}$
corresponding to a flat space with $k=0$, the scale invariant cosmology permits a variety of the  
density parameter $\Omega_{\mathrm{m}}$  (\ref{oprime}) at present for the flat space with $k=0$. 
This is a most interesting property, especially in view 
of the results of the CMB which support a flat Universe \citep{deBern00,Benn03,Planck15}.
However, we have seen in Paper II that for $k=0$ the parameter $\Omega_{\mathrm{m}}$ is necessarily smaller than 1, since
$\Omega_{\lambda} >0 $ and  (\ref{Omegapr}) must be satisfied. 
Expression (\ref{copr}) shows that for  $\Omega_{\mathrm{m}}$ ranging from $0 \rightarrow 1$, $C$ covers 
the range  from $0$ to infinity.
\\

To integrate (\ref{E13}) numerically, we choose a present value for $\Omega_{\mathrm{m}}$, which  determines 
$C$  according to (\ref{copr}) and we proceed to the integration  backwards and forwards in time  starting from the present chosen values $t_0=1$ and $R_0 =1$. The integration provides $R(t)$, its derivatives and the related parameter $H$ and $q$.
Fig. \ref{Rtzero} shows some curves of $R(t)$ for different  $\Omega_{\mathrm{m}} < 1$, all these curves have $k=0$ and 
 $\Omega^*_{\mathrm{m}} =1$. 
Table 1 provides some model data. The value of $H_0$ is given in a scale where $t_0=1$ (column 3), it is also given  (column 8)
in a scale where the time unit  is the age of the Universe $\tau=t_0 -t_{\mathrm{in}}$, while the last column gives the the value of $H_0$  in the current units [km s$^{-1}$ Mpc$^{-1}$].
 To obtain $H_0$ in these  units, we need to have an estimate of th age of the Universe.
 \citet{Frie08} give an estimate of 13.9  $\pm 0.6$  Gyr in a so-called  consensus model  and 13.8 $\pm 0.2$  Gyr in a fiducial model. \citet{Freedman10} provide an age  estimate of 13.7 $\pm 0.5$ Gyr based on three different methods.
The last column of the Table gives the value of $H_0$ expressed in usual units  [km s$^{-1}$ Mpc$^{-1}$] for an adopted age $\tau$ of 13.8 Gyr. This value of the  age of the Universe is also used  in column 7 for obtaining the ages in Gyr.

 In practice to get $H_0$ in [km s$^{-1}$ Mpc$^{-1}$], we proceed in the following way. The inverse of the age of 13.8 Gyr is
  $2.2683 \cdot  10^{-18}$ s$^{-1}$, which in the units currently used for the Hubble constant is equal to 
  70.86 [km s$^{-1}$ Mpc$^{-1}$].  Thus,  this is the value of $H_0$ that  exactly corresponds to $H_0(\tau)=1.000$ in column 8
  of Table 1. On the basis of this correspondence,  we now multiply all values of $H_0(\tau)$ of
   column 8 by 70.86 [km s$^{-1}$ Mpc$^{-1}$]
  to get the values of  $H_0$  in the last column. We see that $H_0=  67$  [km s$^{-1}$ Mpc$^{-1}$] is the Hubble constant predicted 
  for  $\Omega_{\mathrm{m}} =0.30$ in agreement with \citet{Planck15}. The fact that a good agreement
  is obtained  for $\Omega_{\mathrm{m}}=0.30$ indicates that the expansion rate is correctly predicted by the scale invariant 
  models in a consistent way with the age of the Universe. \\


From Table 1 and Fig. \ref{Rtzero}, we note the following properties of the scale invariant models with $k=0$ :
\begin{enumerate}
\item After an initial phase of braking,  there is an acceleration of the expansion, which goes on all the way. 

\item The differences  of the expansion functions $R(t)$ with that of the classical Einstein-de Sitter model (thin broken line in Fig. 1)
are large.

\item No curve $R(t)$ starts with an horizontal tangent, except the case of zero density  which goes like $R(t) \sim t^2 $ (Paper II).

\item All models  with matter start explosively 
with very  high values of $H={\dot{R}} /R$ and a positive value of $q$, indicating  braking.

\item The higher the  input density parameter  $\Omega_{\mathrm{m}}$, the longer the initial braking phase.
 The locations  of the inflexion points  where $q$
changes sign are indicated for the models of different  $\Omega_{\mathrm{m}}$ by a small open circle 
in Fig.  \ref{Rtzero}, see also Table 1. 

\item The lower the density, the longer the present age $\tau= t_0 - t_{\mathrm{in}}$ of the models. This is also true for the
ages given in Gyr.

\item The   properties of the scale factor $\lambda$ deserve some comments. First, we recall that the behavior of $\lambda (t)$
derives from the assumption of the scale invariance of the empty space at macroscopic and large scales. For a totally empty space with 
$\varrho_{\mathrm{m}}=0$, 
the factor $\lambda$ would vary between $\infty$ at the origin, to 1 at present and to zero in an infinite future, 
as shown by the empty model in Paper II.
Fig. \ref{scale} shows that as soon as matter becomes present the amplitude of the $\lambda$-variations falls dramatically.
For example, for a present  $\Omega_{\mathrm{m}}= 0.30$, 
  $\lambda$ varies only from 1.4938 to 1.0 between the origin and the present. Thus, the presence of about  1 H-atom by cubic meter
on the average  is sufficient to shift the initial $\lambda$ value from infinity to about 1.5.
For $\Omega_{\mathrm{m}}$ tending towards unity, the scale factor $\lambda$  tends towards a constant equal to 1. Thus,
the  domain of $\lambda$-values is consistently determined by the matter content or in other words by the departures 
from the scale invariant empty space.

\begin{figure}[t!]
\begin{center}
\includegraphics[width=10.2cm, height=7.0cm]{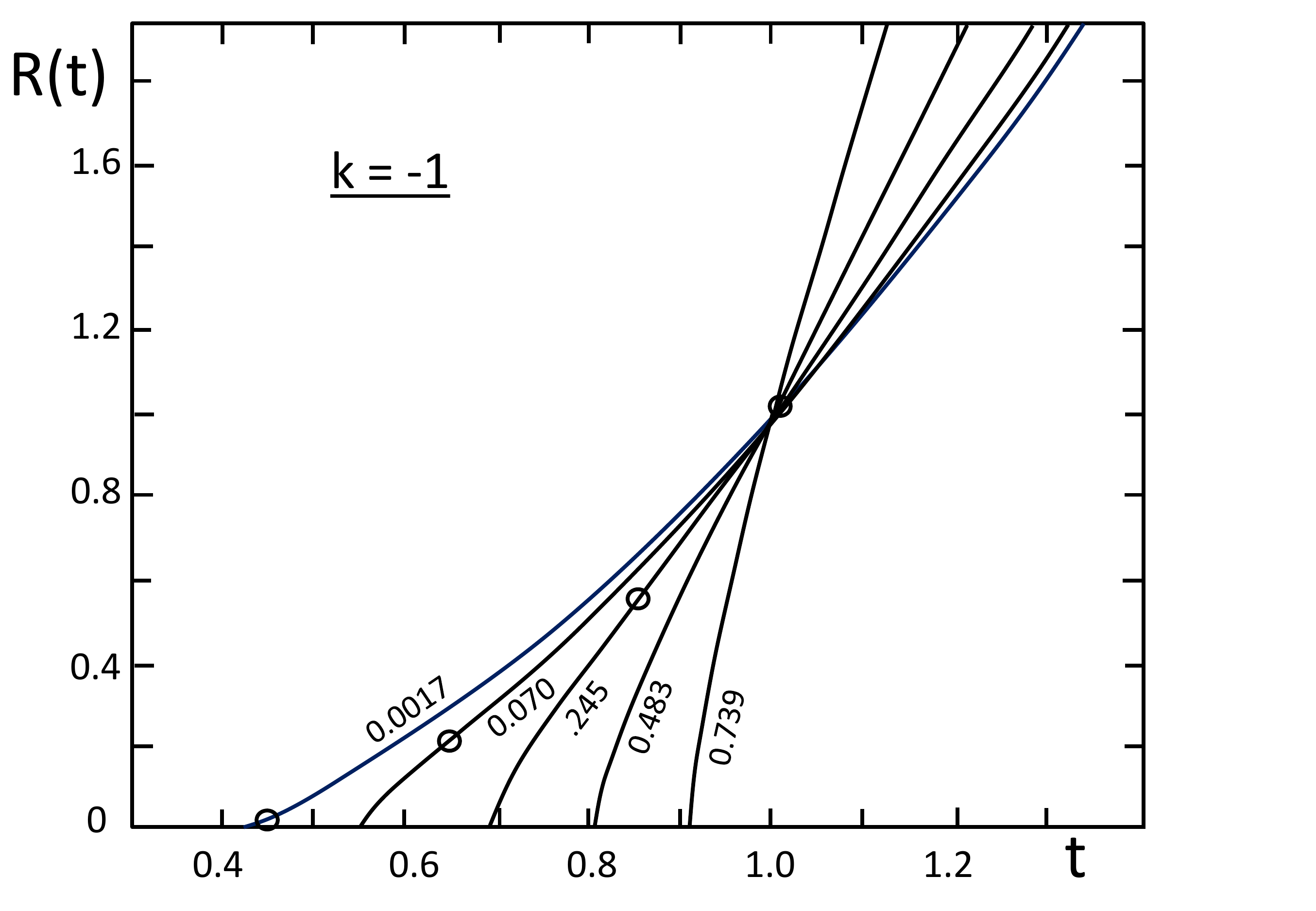}
\caption{ Some solutions of $R(t)$ fro the models with $k=-1$. The curves are labeled by the  values of $\Omega_{\mathrm{m}}$ at $t_0$, the usual density parameter defined by (\ref{oprime}). The corresponding values of $\Omega^*_{\mathrm{m}}$ used to define $C$ 
are 0.001, 0.315, 0.70, 0.90, 0.98 from left to right.}
\label{Rtm1}
\end{center}
\end{figure}

\item The expressions of $q$ are different for the scale invariant and the $\Lambda$CDM models.
For the  flat scale invariant models, $q$ is given at all times by 
\begin{equation}
q = \frac{1}{2} -\Omega_{\lambda} \, ,
\label{qkk}
\end{equation}
\noindent
while for the flat $\Lambda$CDM models, it is
\begin{equation}
q\, = \, \frac{1}{2}\Omega_{\mathrm{m}} -\Omega_{\Lambda} \, .
\label{qlambda}
\end{equation}
\noindent
 The transition from braking to acceleration occurs,  for the flat scale invariant case, when one has the equality
$\Omega_{\mathrm{m}}= \Omega_{\lambda}= 1/2$ at the transition, while in the $\Lambda$CDM model, it occurs when
$\Omega_{\Lambda}= (1/2) \, \Omega_{\mathrm{m}}$, which gives a transition for $\Omega_{\mathrm{m}}=2/3$.

\item  For the flat models with
 $\Omega_{\mathrm{m}}=0.30$, the values of the deceleration parameter $q_0$ at the present time are  
  $q_0=-0.20$ for the scale invariant model and $q_0=-0.55$  for the
 $\Lambda$CDM model.
 The present acceleration  is slightly  stronger in the $\Lambda$CDM 
than in the corresponding scale invariant model.

\item  We  note  the different behaviors of $H_0$ in unit of $t_0=1$   and in unit of $\tau$, the present age of the Universe.
The Hubble constant $H_0$ expressed as a function of the age $\tau$ is smaller for higher densities, the same trend is noted for
$H_0$ expressed in usual units  [km s$^{-1}$ Mpc$^{-1}$].
The particular value  $H_0 = 1/\tau$  is obtained for  $\Omega_{\mathrm{m}} \simeq 0.24$ and $\Omega_{\lambda} \simeq 0.76$
for $k=0$. 

\item  As shown by Table 1,  for the present $\Omega_{\mathrm{m}}= 0.99$, $C$ is equal to 39600 and the model starting at
 $t_{\mathrm{in}}=0.99664$ nearly has a vertical expansion $R(t)$. This  suggests that for $\Omega_{\mathrm{m}}=1$ the
 model   inflates explosively all the way since the orgin. Whether this has some implications at  the origin is an open question.
\end{enumerate}

\begin{figure}[t!]
\begin{center}
\includegraphics[width=10.0cm, height=7.0cm]{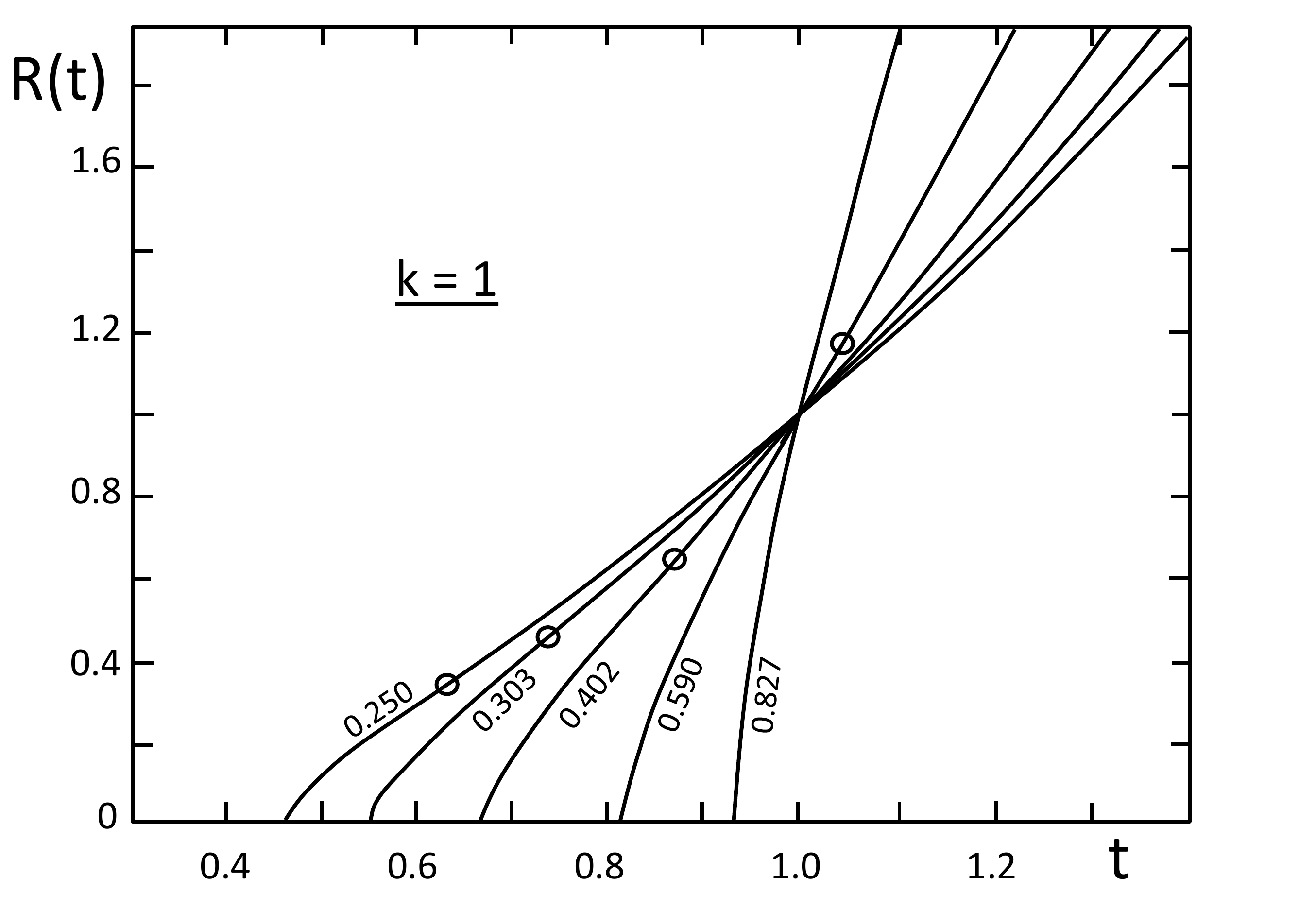}
\caption{Some solutions of $R(t)$ for $k=+1$. The corresponding values of the usual density parameter 
 $\Omega_{\mathrm{m}}$  at $t_0$ are indicated. The values of  $\Omega^*_{\mathrm{m}}$ used to define the $C$-values 
 are 1000, 3.0 , 1.5, 1.10, 1.01. }
\label{Rtk1}
\end{center}
\end{figure}

Below in Table (\ref{detailk0}), we provide the details of the relation $R(t)$ vs. time $t/t_0$,   for  the  density parameter 
$\Omega_{\mathrm{m}}=0.30$, well supported by the \citet{Planck15}. In this table, we also give the redshifts, the corresponding
ages, Hubble parameters and scale factors.

\begin{table*}[t!]  
\vspace*{0mm}
\caption{Cosmological  parameters  of some models with $k=-1$ and $k=+1$ for different values of the density parameter
$\Omega^*_{\mathrm{m}}$ at time $t_0$. The usual density parameter at $t_0$  is $\Omega_{\mathrm{m}}$  in column 8. 
See also the  remarks  for Table 1.}
\begin{center}  \small
\begin{tabular}{ccccccccccccc}
$\Omega^*_{\mathrm{m}}$  &  $C$  & $H_0(t_0)$ & $t_{\mathrm{in}}$ & $q_0$ &
  $\tau$ &  $H_0(\tau)$ & $\Omega_{\mathrm{m}}$ &
$t$(q=0) & $R$(q=0) & $\Omega_{\lambda}$ & $\Omega_{\mathrm{k}}$ &$H_0$ obs  \\
\hline
   &    &   & \\
 & \multicolumn {10} {c}{\bf{k=-1}} \\
0.001  & 0.0010 & 2.4146 & .4157 & -.414 & .5843 & 1.411 & .0002 & 0.424 & 0.009& .828 & 0.172 & 100.0\\
0.100  & 0.1111 & 2.4530 & .4701 & -.398 & .5299 & 1.300 & .019  & 0.531 & 0.104 & .815 & 0.166  & 92.1 \\
0.315  & 0.4599 & 2.5684 & .5467 & -.355 & .4533 & 1.164 & .070  & 0.646 & 0.224 & .779 & 0.152  & 82.5\\
0.500  & 1.0000 & 2.7320 & .6095 & -.299 & .3905 & 1.067 & .134  & 0.734 & 0.342 & .732 & 0.134  & 75.6\\
0.700  & 2.3333 & 3.0817 & .6887 & -.202 & .3113 & 0.959 & .246  & 0.843 & 0.536 & .649 & 0.105 & 68.0 \\
0.900  & 9.0000 & 4.3166 & .8091 & 0.010 & .1909 & 0.824 & .483  & 1.006 & 1.028 & .463 & 0.054 & 58.4\\
0.98   &  49      & 8.1414 & .9090 &  0.247 & .0910 & 0.741 & .739  &  1.142 & 2.076 & .246 & 0.015 & 52.5 \\
0.999 & 999     & 32.639 & .9791  &  0.438 & .0209 & 0.682 & .938  &  1.233& 6.158 & .061  & 0.001 & 48.3\\
\hline
  &    &   & \\
 & \multicolumn {10} {c}{\bf{k=1}} \\
1.001 & 1001    & 32.639 & .9791 & 0.439 & .0209  & 0.682 & .940  & 1.234  & 6.180 & .061 & -.001 & 48.3 \\
1.010 & 101     & 11.050& .9356 & 0.323 &  .0644 & 0.712 & .827  & 1.182  & 2.764 & .181 & -.008  & 50.4\\
1.100 &  11      &  4.3166 & .8157& 0.064 &  .1843 & 0.796 & .590  & 1.042  & 1.179 & .463 & -.054 & 56.4\\
1.5    &   3       &  2.7321 & .6679 &-.164 &  .3321  &0.907 & .402  & 0.872  & 0.657   & .732&-.134  & 64.3 \\
2.0    &   2       &  2.4142 & .6021 & -.243&  .3979  &0.961 & .343  & 0.797  & 0.534  & .828 & -.172 & 68.1 \\
3.0    &  1.5     &  2.2247 & .5475 & -.298 &  .4525 & 1.007 & .303 & 0.736  & 0.455  & .899 & -.202  & 71.3\\
10.0  & 1.1111 &  2.0541  & .4819 & -.355 & .5181 & 1.064 & .263 & 0.662  & 0.379 &  .974 & -.237 & 75.4\\
1000 & 1.0001 &  2.0005  & .4564 & -.375 & .5436  & 1.088 & .250 & 0.634 & 0.354 & 1.00 & -.250 & 77.0 \\
\hline 
\normalsize
\end{tabular}
\end{center}
\end{table*}

\section{The elliptic and hyperbolic scale invariant models}

Although the non-Euclidean models  are not supported by the observations of the CMB radiation \citep{Planck15},
 we briefly present the main 
properties of these models. 
We first have to relate the constant $C$ to    the density  parameters. Expressing $C$ with (\ref{E12})
and (\ref{roc}), we get at time $t_0$

\begin{equation}
C= \frac{8 \, \pi \,  G  \varrho_{\mathrm{m}}}{3} = \Omega^*_{\mathrm{m}} \, H^2_0 \, \left(1-\frac{2}{t_0 H_0} \right)  \,  ,
\label{coc}
\end{equation}
\noindent
and with (\ref{deuxo})

\begin{equation}
\mathrm{and} \quad C \,=  \,\Omega_{\mathrm{m}} \, H^2_0 \, .
\label{coco}
\end{equation}

\noindent 
We see  that the real density $\varrho_{\mathrm{m}}$ at the present time behaves
like $C$ and thus as $\Omega_{\mathrm{m}}  \, H^2_0$.
From the basic equation (\ref{E1})  and the definition   (\ref{roc}) of the critical density 
$\varrho^*_{\mathrm{c}}$,  we also have  the following relation between the geometrical parameter $k$ 
and  $\Omega^*_{\mathrm{m}}$  at the present time,

\begin{equation}
\frac{k}{R^2_0} =H^2 _0\left[ (\Omega^*_{\mathrm{m}} -1) \left(1 - \frac{2}{t_0 H_0} \right )\right]  \, , 
\label{kom}
\end{equation}
\noindent 
which was relation (40) of Paper II. It allows us to eliminate $[1-2/(t_0 \,H_0)]$  from (\ref{coc}) and  obtain 

\begin{equation}
C \, = \,  \frac{k \, \Omega^*_{\mathrm{m}}}{\Omega^*_{\mathrm{m}}-1}  \, , \quad \mathrm{with} \; \; k= \pm 1  \, .                            
\label{Ck1}
\end{equation}
\noindent
A model is defined by its $\Omega^*_{\mathrm{m}} $-value at the present time.
For integrating equation (\ref{E12}),  we first choose an arbitrary value of  $\Omega^*_{\mathrm{m}}$  for the considered $k$
and then use   (\ref{Ck1})  to obtain the  corresponding $C$--value.  The integration of  (\ref{E12}) from the present  
$t_0 =1$  and $R_0 =1$ is performed forwards and backwards to obtain $R(t)$   and 
its first and second derivatives. The value of $H_0=(\dot{R}/R)_0$  at the present time
gives us the    $\Omega_{\mathrm{m}}$-value corresponding to the chosen $\Omega^*_{\mathrm{m}}$,
 according to relation  (\ref{deuxo}).
 
 Here, for non zero curvature models, $\Omega_{\mathrm{m}} \neq  (1-\Omega_{\lambda})$ at all times and
 we do not have $\Omega^*_{\mathrm{m}}$ equal to 1 as for $k=0$.  $\Omega_{\mathrm{m}}$, 
 $\Omega_{\mathrm{k}}$ and $\Omega_{\lambda}$, as well as $\Omega^*_{\mathrm{m}}$
 vary with time in these models. We have seen in Sect. 4.1 of Paper II, that for $k=-1$, the variety of 
 scale invariant models is necessarily restricted to those with $\Omega_{\mathrm{m}} < 1$. For $k=1$, we found that if 
 the condition $(\Omega_{\mathrm{k}}+ \Omega_{\lambda}) > 0$ is satisfied, the variety of models is also
  restricted to those with $\Omega_{\mathrm{m}} <1$. From Table 2, we see that this condition is satisfied, this is why
  both sets of models with $k= \pm 1$ have the usual density parameter  $\Omega_{\mathrm{m}} < 1$.

\begin{figure}[t]
\begin{center}
\includegraphics[width=10.1cm, height=7.5cm]{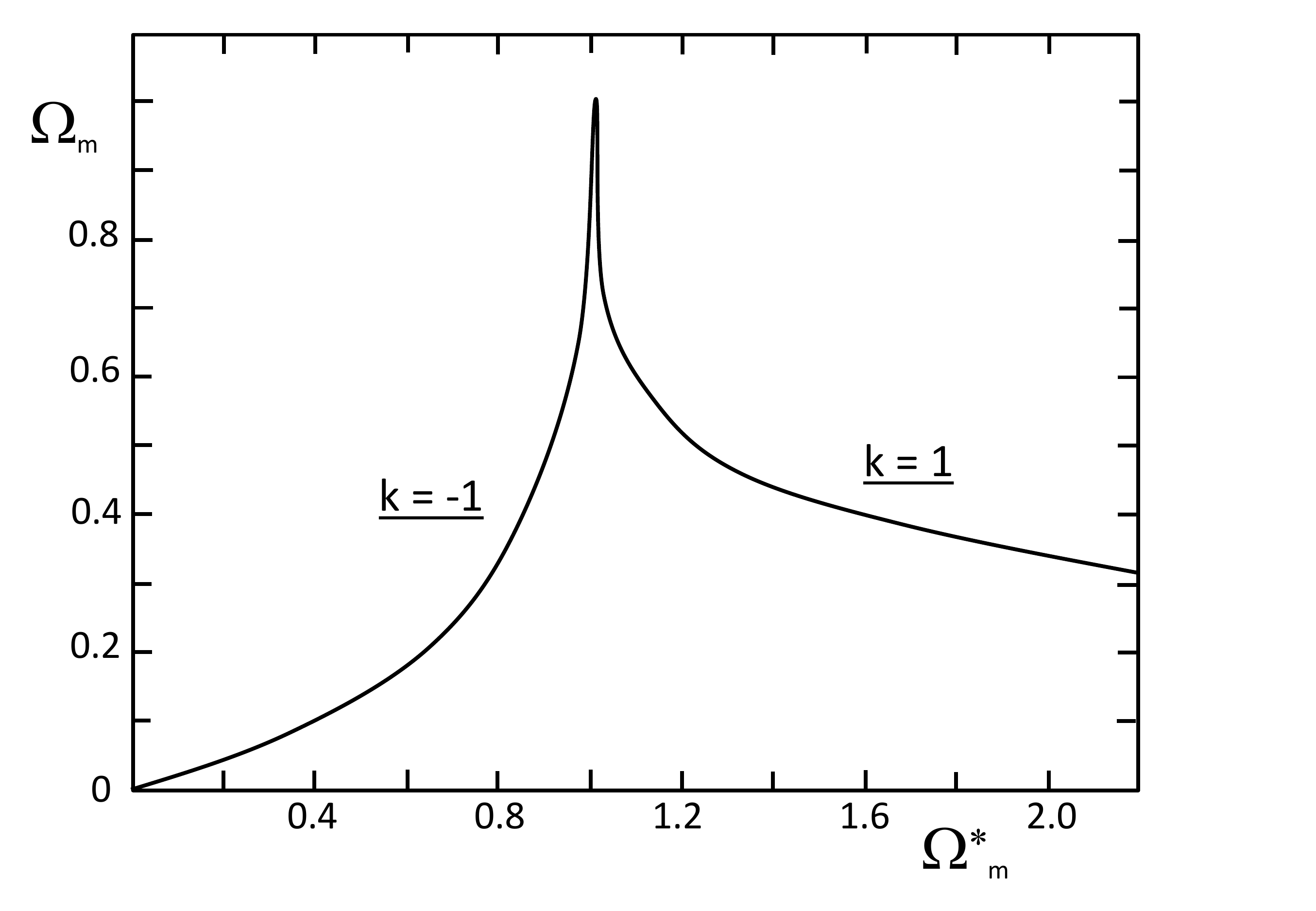}
\caption{Relation between the two density parameters defined by (\ref{roc}) and (\ref{oprime}) at time $t_0$.  $\Omega_{\mathrm{m}}$ is the usual density parameter at present. For large values of $\Omega^*_{\mathrm{m}}$, the values of of $\Omega_{\mathrm{m}}$ tend towards an asymptotic limit of 0.25.}
\label{DeuxOm}
\end{center}
\end{figure}

Figs. \ref{Rtm1} and \ref{Rtk1} illustrate some solutions for $k= \pm 1$ and Table 2
gives some model parameters for different values of  $\Omega_{\mathrm{m}}$.
 From these figures,  we see that the three families 
of $R(t)$ curves for $k=0$ and $k=\pm 1$ are on the whole not so different from each other.
 The curves for $k=\pm 1$ also show the same succession with first a braking  and then an acceleration phase.
For lower $\Omega_{\mathrm{m}}$, the initial expansion is less steep and starts earlier, while for $\Omega_{\mathrm{m}}$
approaching 1 the expansion tends to become explosive, as already seen for $k=0$.
  The relative similarity of the three families of curves  indicates that the curvature
   term $k$ has a limited effect compared to the  density (expressed by $C$ in the equations)  and to 
the acceleration  resulting from scale invariance. Unlike the Friedman models, the same density parameters $\Omega_{\mathrm{m}}$
may exist for different curvatures.

As for models with $k=0$, the models with $k= \pm1$ may have all possible values of $C$, and thus of $\varrho_{\mathrm{m}}$, from 0 to infinity.
However,  they all have the usual density parameter 
$\Omega_{\mathrm{m}}$ smaller than 1.0, as mentioned above. For $k=-1$,  the two density parameters $\Omega_{\mathrm{m}}$ and
$\Omega^*_{\mathrm{m}}$  cover the range from 0 to 1, which   
is not particular. 
 However, for $k=+1$, the behavior of the parameters is peculiar, as illustrated by Table 2. When $C$ increases from 1 to infinity, 
 $\varrho$ increases from a minimum value  to infinity. At the same time,
  $\Omega^*_{\mathrm{m}}$ decreases from infinity to 1.0, while 
 $\Omega_{\mathrm{m}}$ goes from a limit of 0.25 to 1.0.

Fig. \ref{DeuxOm} illustrates the relation between the two density parameters at $t_0$, 
$\Omega_{\mathrm{m}} =\Omega^*_{\mathrm{m}}[1 -2/(t_0  \,H_0)]$.
 For $k=-1$, $\Omega_{\mathrm{m}}$ grows first much slower  than $\Omega^*_{\mathrm{m}}$ 
 due to the subtraction of the term $2/(t_0  \,H_0)$.
Then as $H_0$ becomes very large, $\Omega_{\mathrm{m}}$ grows fast. 
For $k=1$, as $\Omega^*_{\mathrm{m}}$  increases we have the opposite for the usual density parameter, this results
from the fact that the term   $[1-2/(t_0 \, H_0)]$  becomes very small. As an example from Table 2,
 for $\Omega^*_{\mathrm{m}}=1000$,  $H_0= 2.00050$, so that the term $[1 -2/(t_0  \,H_0)] =0.00025$ and
 $\Omega_{\mathrm{m}}=0.25$.   In all comparisons with observations, we will evidently use the
  $\Omega_{\mathrm{m}}$-parameter.

\section{Comparisons of models and observations: the density parameters and the Hubble constant at present}  \label{compobs}

\begin{figure}[t!]
\begin{center}
\includegraphics[width=9.8cm, height=7.8cm]{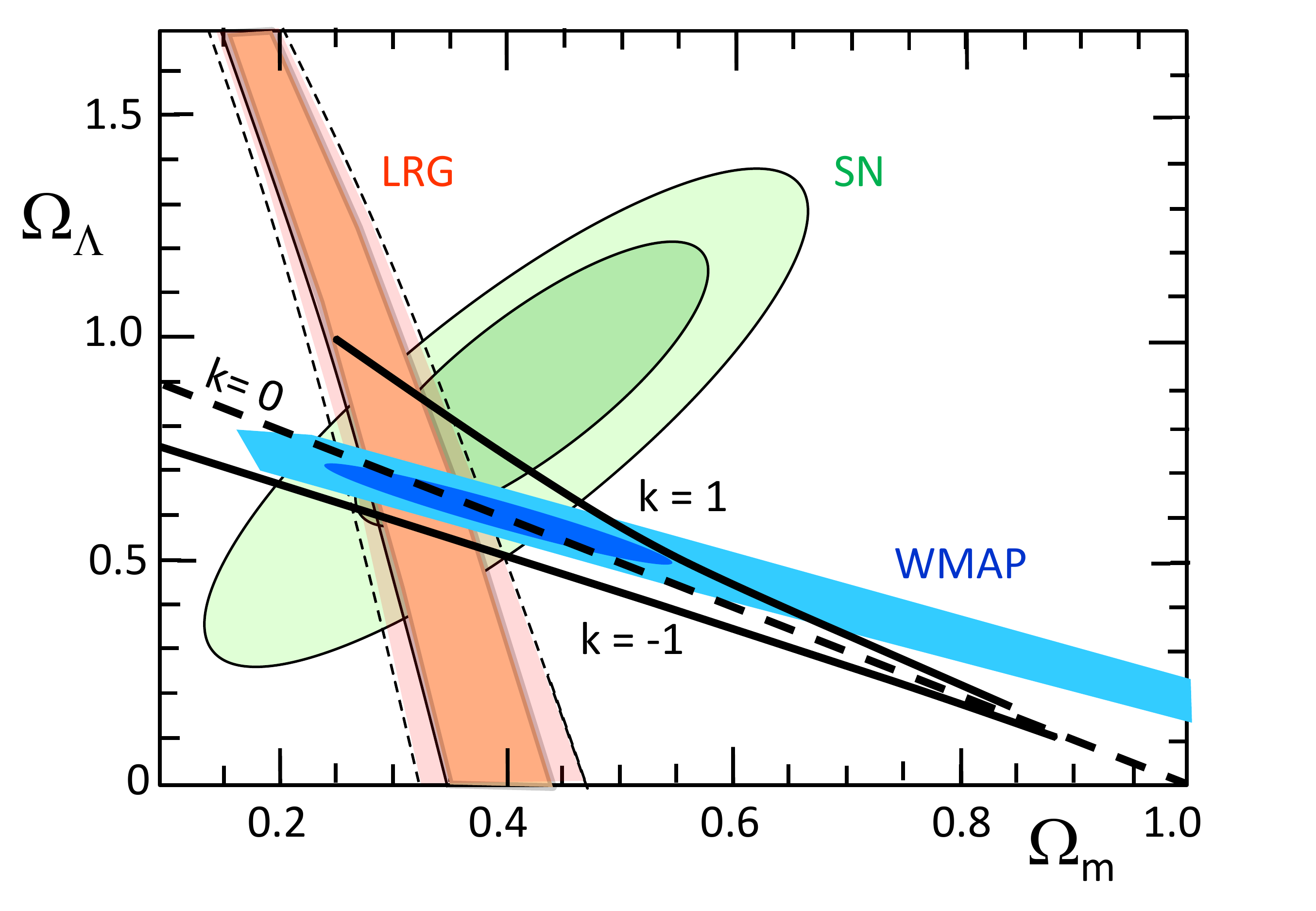}
\caption{The constraints on  $\Omega_{\Lambda}$ vs. the present $\Omega_{\mathrm{m}}$ from the observational data collected by \citet{Reid10},  with the
constraints from WMAP5, from the Union SN sample and from 
the halo density field of luminous red galaxies of the SDSS DR7
as analyzed by \citet{Reid10}. We have superposed  the  results of the scale invariant models
for $\Omega_{\lambda}$  and  $\Omega_{\mathrm{m}}$ (at $t_0$) from Tables 1 and 2 for the different curvature parameters  $k$.}
\label{OmegaReid}
\end{center}
\end{figure}

\begin{figure*}[t!]
\centering
\includegraphics[width=.85\textwidth]{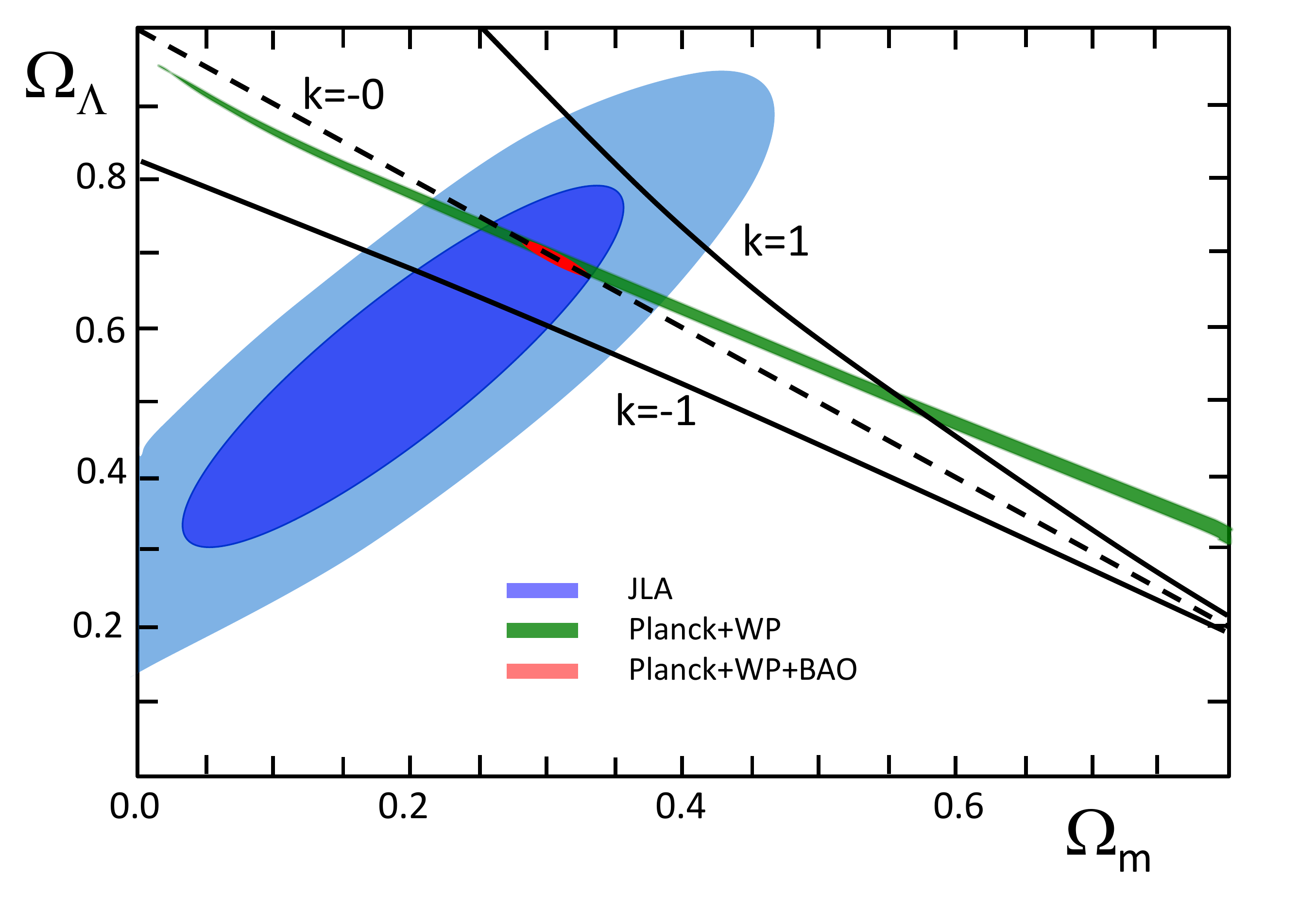}
\caption{The  $\Omega_{\Lambda}$ vs. the usual $\Omega_{\mathrm{m}}$ from  the observational data  collected by \citet{Betou14}, cf. their Fig. 15. The SN sample from JLA  (blue) is superposed with the Planck temperature and WMAP polarization measurements
(green). The most stringent constraint (red) accounts for 
the BAO results. 
   We have superposed  the scale invariant model results 
for $\Omega_{\lambda}$  and  $\Omega_{\mathrm{m}}$ from Tables 1 and 2 for the different curvature parameters.}
\label{Betou}
\end{figure*}

Comparisons with  observations are essential to invalidate or validate theories. 
In this section, we make  comparisons for several important properties, in particular the density parameters and the expansion rate $H_0$.

\subsection{The $\Omega$--parameters}

Since the discovery of the acceleration of the expansion, a number of constraints on the $\Omega$--parameters
have been found and analyzed in recent major works.  The studies of the CMB with Boomerang  \citep{deBern00}, WMAP \citep{Benn03} and the  \citet{Planck15} support  more and more the flatness  $k=0$  of the Universe.  For example,
 the last Planck results \citep{Planck15} give a value $\Omega_{\mathrm{k}}= 0.00 \pm 0.005$ at a 95\% confidence limit.
Over recent years,  the various surveys globally converge  towards  similar results within always more stringent limits.

\citet{Frie08} found average  values of $\Omega_{\mathrm{m}}= 0.246 \pm0.028$ and $\Omega_{\Lambda}=0.757 \pm 0.021$,
 their reference study was based on the magnitude-redshift data for supernovae, the CMB radiation
measured by WMAP, the age constraints and the baryon acoustic oscillations (BAO).  In this technique, one considers 
 that the initial oscillations  in the CMB, with a lengthscale
defined by the sound velocity in the plasma,  influence the clustering of galaxies  and provide a reference length scale (150 Mpc), 
which is used to measure the cosmic distances and probe the acceleration of expansion. 
The  analysis of the BAO from a sample of 893'319 galaxies in the  Sloan Digital Sky Survey (SDSS)  Data Release 7 (DR7) by \citet{Percival10} leads to a slightly higher density ($\Omega_{\mathrm{m}}= 0.286 \pm0.018$). 
 \citet{Reid10} examine the constraints from the clustering of luminous red galaxies
in the SDSS DR7.  The power spectrum of the halo  density field of galaxies is sensitive to
 the dark matter density $\Omega_{\Lambda}$. Combining their data with WMAP 5 years results, they find
$\Omega_{\mathrm{m}}= 0.289 \pm0.019$, ($\Omega_{\Lambda}$ is here the complement to an $\Omega$-sum of 1.011$\pm0.009$).

Clusters of galaxies provide another interesting constraint on the density parameters \citep{Allen11}. Let $f_{\mathrm{gas}}$ 
be the ratio of the mass in the form of X-ray emitting gas to the total mass in clusters. This ratio in the largest concentrations of mass in the Universe is generally
 assumed constant and about equal to the baryon fraction. The assumption of  a constant $f_{\mathrm{gas}}$ with redshift $z$ places constraints on the cosmological models. Combining these constraint  with those of the CMB and supernovae leads  to  $\Omega_{\mathrm{m}}= 0.275 \pm0.015$ and $\Omega_{\Lambda}=0.725 \pm 0.016$. A recent study by \citet{Betou14} of the cosmological  parameters  with the  
project Joint Light-curve Analysis (JLA) combines the 
supernova results of two major surveys the SDSS and SNLS (SN Legacy Survey) together with the CMB data from Planck and 
WMAP, including also the constraints from BAO. This study  gives very 
stringent conditions as illustrated by Fig. \ref{Betou} and favors
a value $\Omega_{\mathrm{m}}= 0.295 \pm0.034$ .

Fig. \ref{OmegaReid} based on the results by \citet{Reid10} and Fig.  \ref{Betou} based on the  recent and very constraining results by \citet{Betou14} show  the comparison of the observed density parameters $\Omega_{\mathrm{m}}$
and $\Omega_{\Lambda }$ with the results  of our models. In the scale invariant models, $\Omega_{\lambda}$  represents
the contribution of the effects of scale invariance to the energy-density. 
 The flat  model with  $k=0$  and  $\Omega_{\mathrm{m}} \approx 0.30 $ remarkably well fits the various constraints. The two sets of models with non-zero curvature do not agree with observations, particularly the models with $k=1$.

\begin{figure}[h]
\begin{center}
\includegraphics[width=11.0cm, height=7.5cm]{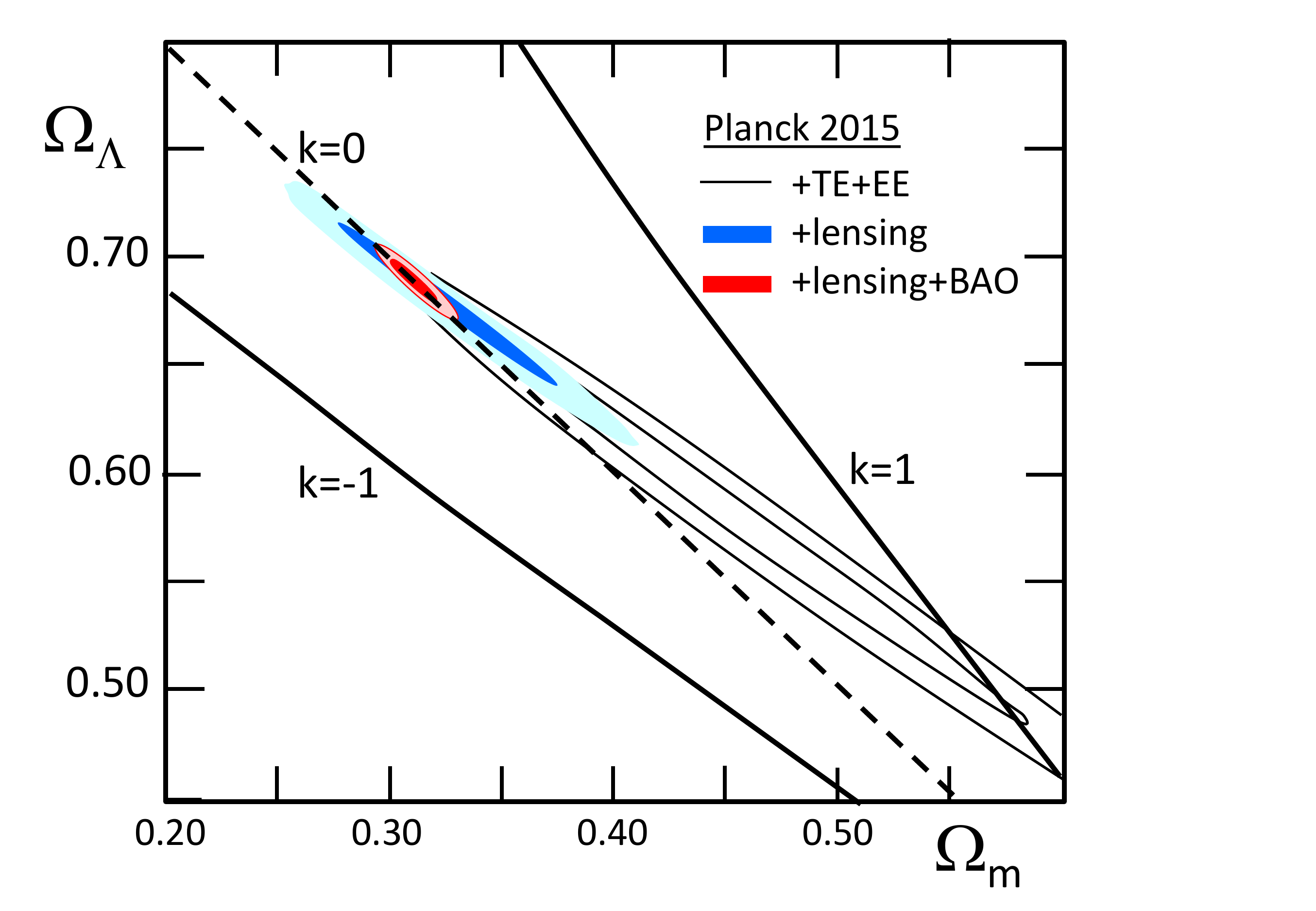}
\caption{Data from Fig. 26 in  \citet{Planck15}: the black contours are 
from the various Planck spectra. The account of the lensing effect on temperature and polarization tightens the permitted domain (blue contours). The inclusion of both lensing and BAO constraints dramatically restrains the permitted domain (red contours). The thick black lines show the scale invariant model results as in Fig. \ref{Betou}.
}
\label{OmegaPlanck15}
\end{center}
\end{figure}

The successive releases of CMB data from Boomerang, WMAP and Planck more and more constrain  the density parameters. 
The Planck data particularly when combined with the BAO  tightens very much the permitted interval for the $\Omega$-values. 
The  Planck results \citep{Planck15} support   $\Omega_{\mathrm{m}}=0.308 \pm 0.012$  in a flat model Universe. Fig. \ref{OmegaPlanck15} compares these last results with the various models.
We notice the strongly constrained red zone and its perfect agreement with the $k=0$ scale invariant models for the above value
of  $\Omega_{\mathrm{m}}$.

This confirms that a scale invariant model correctly account for the observed matter density and  
acceleration of the expansion, or in other words for the amount of the supposed dark matter.
Thus, as far as the  density parameters are concerned, the scale invariant cosmology shows agreement with observations.
These results are  encouraging to pursue the exploration of the consequences of the scale invariant cosmology.

\subsection{The Hubble constant $H_0$ in relation with the $\Omega$-parameters}  \label{Hh}

Another important test  concerns the value of the Hubble constant at the present time $H_0$. The models 
 internally provide the Hubble constant $H_0(\tau)$ as a  function of the present age $\tau$  of the Universe (e.g. column 8 in Table 1). As seen above  to get the value 
of $H_0$ in [km s$^{-1}$ Mpc$^{-1}$] from the models, we need both  the present expansion rate $H_0(\tau)$ given by the models and
 an estimate of the present age of the Universe. 
In  Tables 1 and 2, we have adopted 
 an age of 13.8 Gyr consistent with the best present estimates and to derive the $H_0$-values corresponding
 to different parameters  we proceed as explained in Sect. \ref{flat}.

There has always been scatter in the results for  $H_0$, this is still the case at present, although it is now much decreasing. 
\citet{Frie08} give a value $H_0= 72 \pm 5 $ in [km s$^{-1}$ Mpc$^{-1}$],  $73 \pm 4$ is obtained 
by \citet{Freedman10}, $68.2 \pm 2.2$ by \citet{Percival10},
  $69.4 \pm 1.6$ by \citet{Reid10}, $70.2 \pm 1.4$ by \citet{Allen11}, $67.8 \pm 0.9$ by the \citet{Planck15}.

The models in Table 1 for $k=0$ show the dependence of $H_0$ on the matter density. $H_0$ expressed in current units
 consistently 
decreases for an increasing matter density, since braking is more efficient.  For values between 
$\Omega_{\mathrm{m}}=0.246$ and 0.308 corresponding to the values given by \citet{Frie08} and the \citet{Planck15},
we get values of $H_0$ between 70.2 and  66.5    [km s$^{-1}$ Mpc$^{-1}$], a range very consistent with the observed one.
If we would have adopted an age of 13.7 Gyr, these values would have been 67.0 and 70.7  and for an age of 13.9 Gyr,
66.0 and 69.7 respectively, values which would not change the conclusions.

\begin{figure}[t!]
\begin{center}
\includegraphics[width=11.4cm, height=7.5cm]{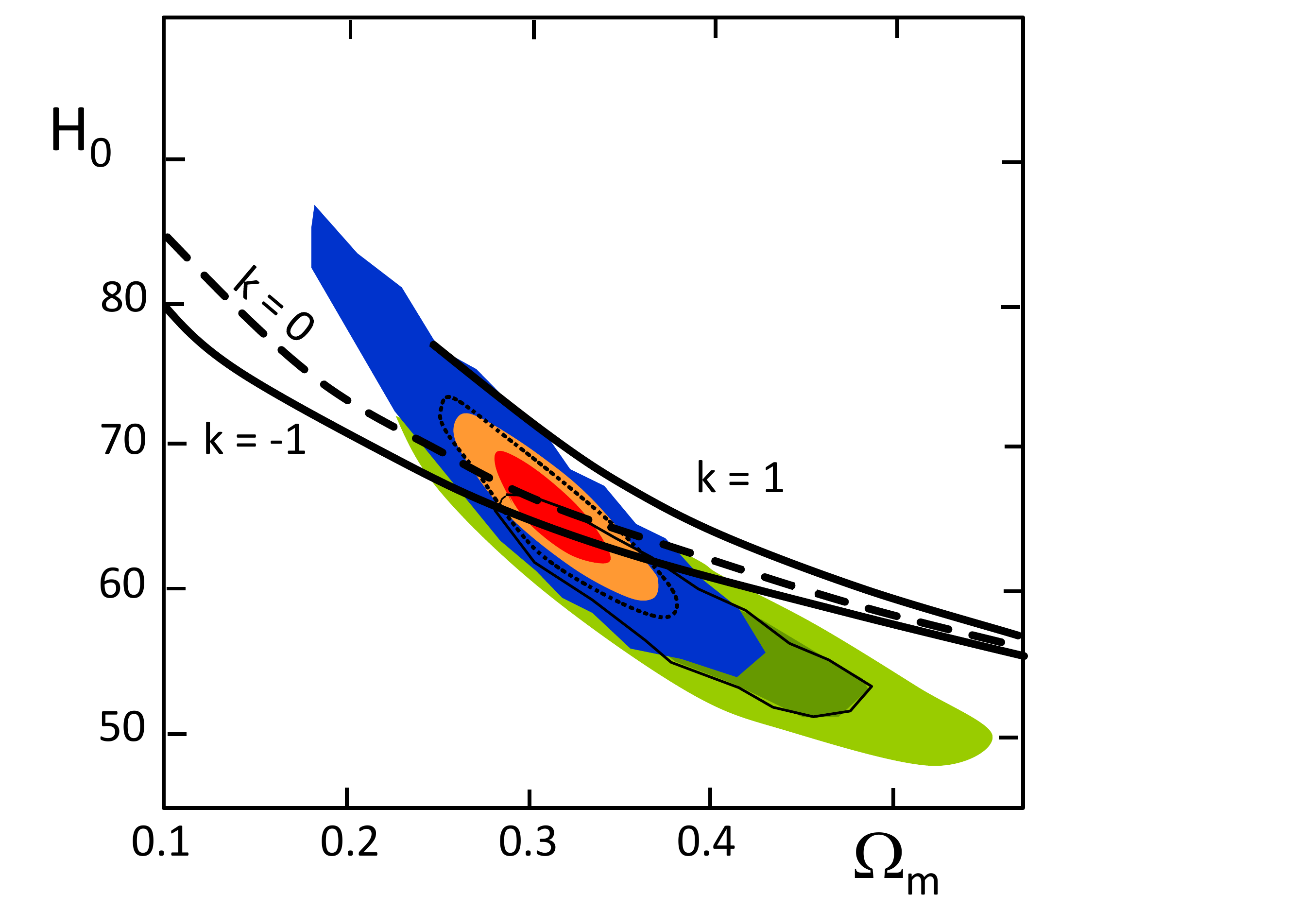}
\caption{Observational data from Fig. 10 by \citet{Reid10}.
The green contours represents the constraints from WMAP+SN,
the blue from WMAP5+luminous red galaxies (LRG), the orange and red from WMAP5+SN+LRG in the $\Lambda$CDM models with free curvature and constant $w$-parameter for the equation of state.
The scale invariant models of Table 1 and 2 with different curvature are over-plotted (black lines).
}
\label{Hzerom}
\end{center}
\end{figure}

 Fig. \ref{Hzerom} present the constraints on the $H_0$ values vs.
 the density parameter $\Omega_{\mathrm{m}}$ derived from the CMB, SN and clustering of LRG  within the CDM models
with free curvature and a constant $w$-parameter \citep{Reid10}. 
Such a comparison is testing whether the present expansion rate $\dot{R}(t_0)/R(t_0)$ predicted by   the models for the observed  matter density $\Omega_{\mathrm{m}}$
is consistent with observations. 

We see that the curve defined by the $k=0$ models nicely fits the central red zone, best constrained 
by the WMAP5 data together and the results from the clustering. The scale invariant models with $k= -1$
are not so much different from those with $k=0$, while the models with $k=1$
do not agree with the observational constraints. We may also do the comparison with the recent Planck
data. For a matter density of   $\Omega_{\mathrm{m}}=0.308 \pm 0.012$, a value of $H_0= 67.8 \pm 0.9$
[km s$^{-1}$ Mpc$^{-1}$] is obtained in the $\Lambda$CDM model by the Planck collaboration. The scale invariant model with $k=0$
gives for the above  density 66.5 $\pm0.7$. Thus, we
note that the agreement for the constraints set by $H_0$ vs. matter density is quite good.

Fig. \ref{omkH} compares models and data from the \citet{Planck15} in the $\Omega_{\mathrm{k}}$ vs. $H_0$ plot. 
We verify that the models with $k=0$ perfectly cross the region defined by the Planck  and BAO constraints,
 for a value $H_0$  very well corresponding to the  Planck results.
In this plot, the models with $k = \pm 1$ strongly diverge from  observations.

On the whole, the scale invariant cosmological models give in Table 1 a value of the Hubble constant $H_0$ 
 in agreement with observation for $\Omega_{\mathrm{m}}=0.30$.
 The plots of $H_0$ vs. the density parameters $\Omega_{\mathrm{m}}$
and $\Omega_{\mathrm{k}}$ show an excellent agreement for a scale invariant model with
$k=0$ and  $\Omega_{\mathrm{m}} \approx 0.30$. 

The tests we have made above concern the model properties at the present time. We have performed comparisons of the predictions of the scale invariant models with the recent observational constraints from the SN Ia, the BAO oscillations and CMB data concerning the  energy-density parameters $\Omega_{\mathrm{m}}$, $\Omega_{\mathrm{k}}$ and $\Omega_{\Lambda}$, we have also examined the present expansion rate $H_0$  and its relation to the energy-density parameters.
We now turn to some
tests concerning  different epochs in the evolution of the Universe.

\section{Observational dynamical tests at other epochs}

A major prediction of the cosmological models, including the scale invariant models, concerns the expansion
history $R(t)$ of the Universe.  The results      depend on the basic equations with the 
conservation laws implied by the model equations. The tests we now perform concern  past epochs in the
  history of the Universe. Several   observational tests on the past dynamics of the Universe were successfully developed over the last decades.
We may mention among  others:

\begin{figure}[t!]
\begin{center}
\includegraphics[width=12.6cm, height=7.6cm]{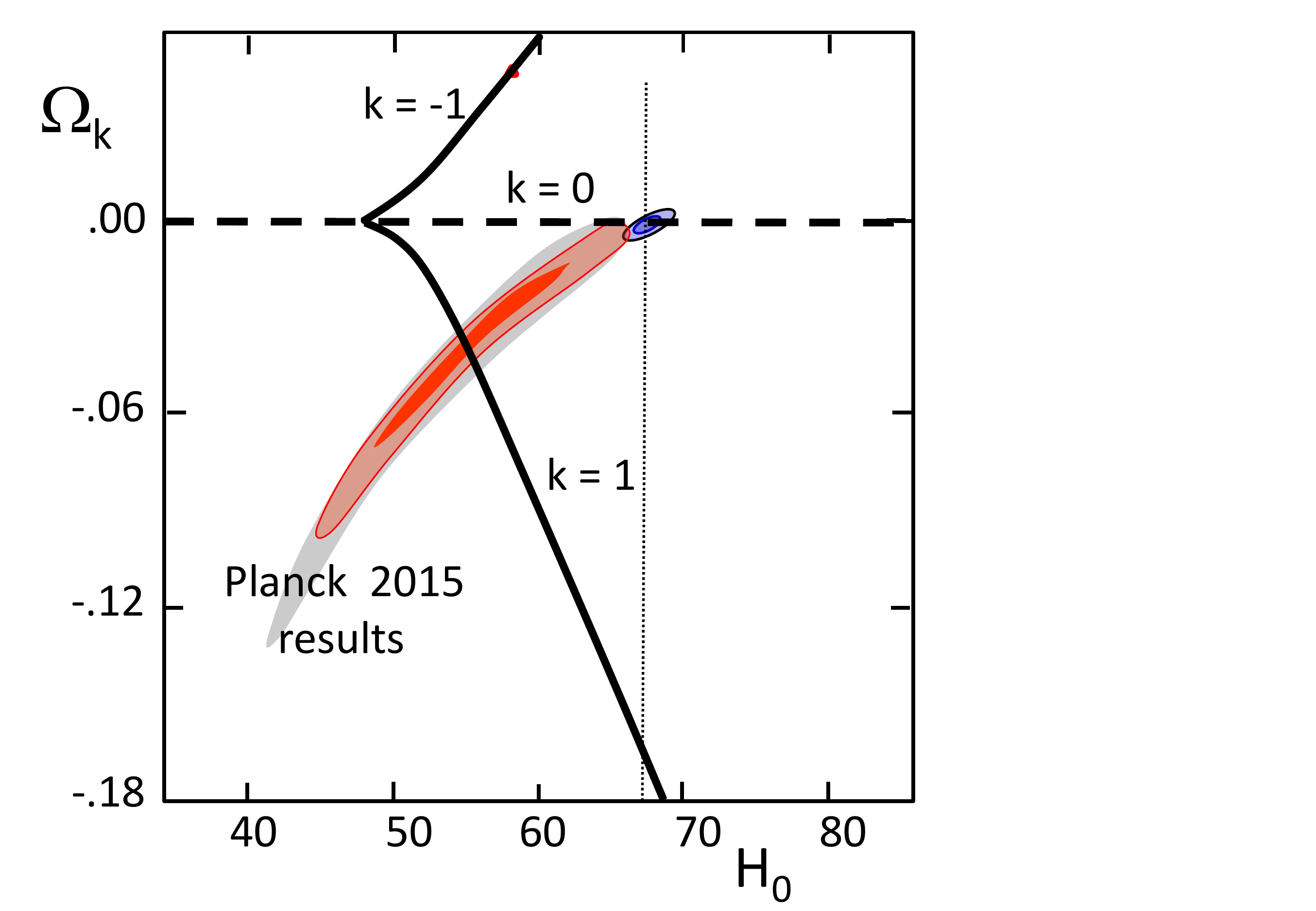}
\caption{Data from Fig. 26 in  \citet{Planck15}, the gray zone only uses the TT spectra and polarization data, the red zone applies
all spectra and polarization results, while the blue accounts in addition for the BAO constraints. The scale invariant  models with different 
curvature $k$ are shown by thick black lines.
}
\label{omkH}
\end{center}
\end{figure}

- The Hubble or magnitude-redshit (m-z) diagram based on distant supernovae of type Ia used as standard candles  \citep{Riess98,Perl99}.

- The preferred 
length-scale given by BAO provides a standard of length at large distances. The BAO may be observed in large galaxy and quasar surveys \citep{Eisen05}. 

- In the case of a very large survey, the preferred scale from BAO and large clusters may be 
studied in both the radial and tangential directions under the assumption that the observed objects
are isotropic. This method first devised by \citet{Alcock79} allows one to test cosmological models, giving for example
indications on both the angular distance and on the expansion rate
$H(z)$ at the considered redshift,  see also \citet{Blake12,Busca13}. 

- The method of ''cosmic chronometers'' is based on the simple relation
\begin{equation}
H(z) =- \frac{1}{1+z} \, \frac{dz}{dt} \, ,
\end{equation}
\noindent
obtained from $R_0/R=1+z$ and the definition of $H=\dot{R}/R$. The critical
ratio $dz/dt$ is estimated from of a sample of passive galaxies 
(with ideally no active star formation) of different redshifts and age estimates \citep{Jim02,Simon05,Melia15,Moresco15}.

\begin{figure*}[t!]
\centering
\includegraphics[width=.95\textwidth]{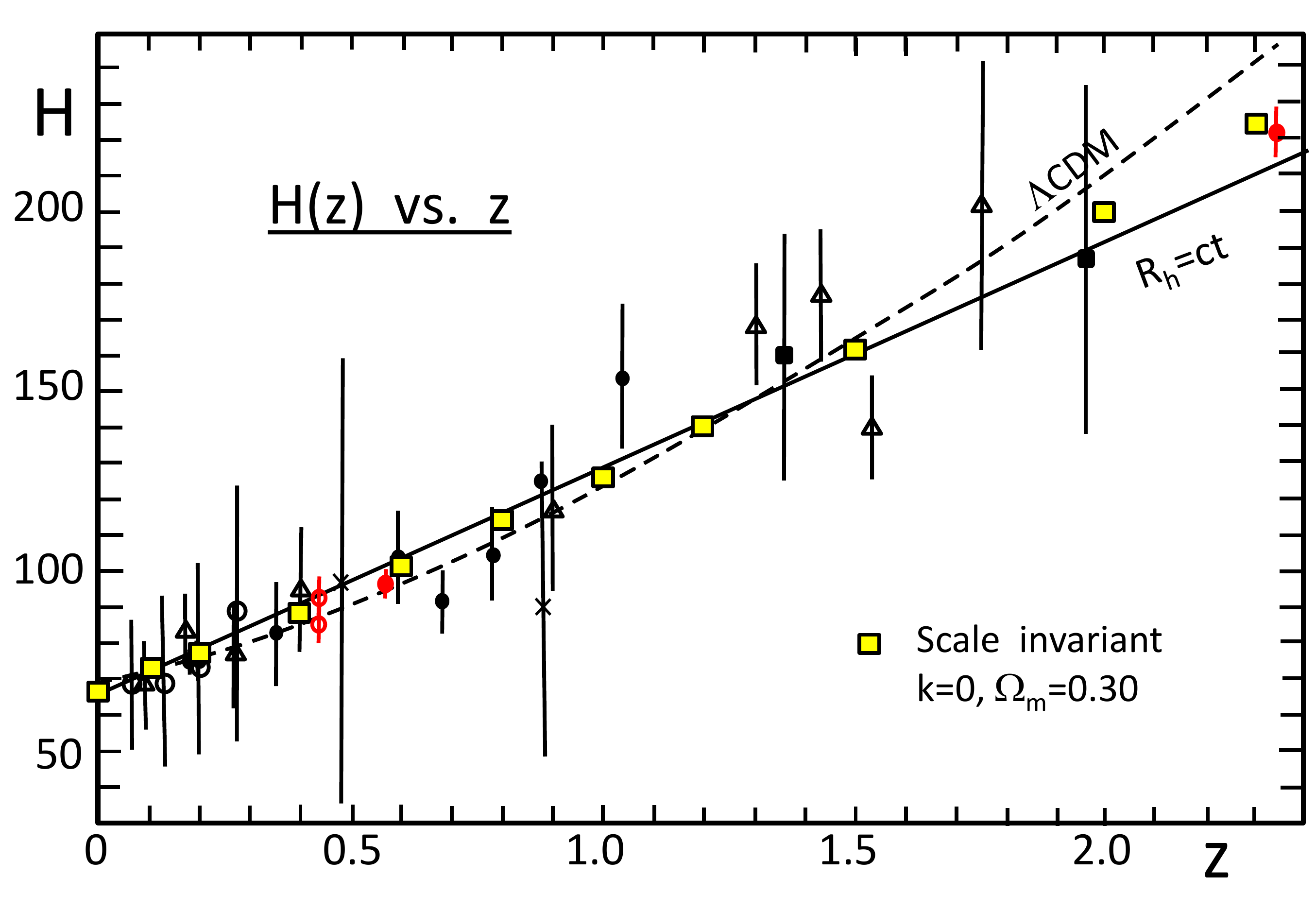}
\caption{The $H(z)$ vs. redshift plot, with $H$ in km s$^{-1}$ Mpc$^{-1}$.
The observations are the model-independent data collected by \citet{Melia15}. The black square are from \citet{Moresco15}, the open circles from \citet{Zhang14}, the black circles from \citet{Moresco12}, the crosses by \citet{Stern10}, the open triangles from \citet{Simon05}. Three other recent  model independent and high precision  data are added as red points.The filled red point at z=2.34 is from the BAO data of the BOSS DR11 quasars by \citet{Delubac15}, the filled red point at z=0.57 is from the BAO given 
by clustering of galaxies in the BOSS from SDSS-III by
 \citet{Anderson14}, the two open and connected red points at z=0.43 represent one sample from an analysis by \citet{Moresco16} of data from BOSS DR9  interpreted with two different sets of models of spectral evolution.
 The curves for the $\Lambda$CDM and $R_h=c \,t$ models are by \citet{Melia15}. The yellow squares indicate the predictions of the scale invariant model for $k=0$ and $\Omega_{\mathrm{m}}=0.30$.  }
\label{history}
\end{figure*}

\subsection{The expansion history of  the Universe}

The determination of the expansion rate $H(z)$ vs. redshift represents a 
 direct and constraining  test on the expansion function $R(t)$ over the ages.
In order to perform valid tests of the cosmological models, it is essential that the observational data are  independent on the cosmological
models, otherwise the results may
be biased towards the used model.  The method of the cosmic chronometer appears as a powerful one, since
 there is no assumption depending on a particular cosmological  model,  as emphasized   by several  authors, namely \citet{Simon05,Stern10,Melia15, Moresco15,Moresco16}.  In several  cases, as pointed out by  \citet{Melia15}, ''cosmological observations'' based solely on BAO may have some  dependence   on the tested cosmology. We note, however, that the method of cosmic chronometers, although independent on the cosmological models, depends on the models of spectral evolution of galaxies, which are mainly based on the theory of stellar evoluton.
 This illustrates the well known fact that all cosmological tests have their weak and strong points.
 
 Table \ref{detailk0} shows many properties of the scale invariant model with $k=0$ and $\Omega_{\mathrm{m}}=0.30$ as functions of redshift $z$.
 Column 7 gives the Hubble values $H(z)$ for different redshifts. These values of $H(z)$ are derived in the same way as for Table 1.
 To perform comparisons between models and observations, we use the data by \citet{Simon05,Stern10,Moresco12,Zhang14,Moresco15} as collected by \citet{Melia15}, completed by other recent high precision and model independent data (shown in red colour) by
 \citet{Anderson14,Delubac15,Moresco16}. Fig. \ref{history} presents these
 data with different symbols according to the authors. The two connected open red circles at $z=0.43$ 
 concern the same BAO at $z=0.43$, 
 but where the ages are based on two different 
 models of evolving passive galaxies \citep{Moresco16}. We see that, at least here,
  the differences due to different models of stellar populations are rather limited. In this figure, we have also reported the $\Lambda$CDM model and a  model
where $R(t)$ linearly increases with time like the horizon $R_h=c \, t$ \citep{Melia15}. According to these authors, this last model is
 better supported by different  observations  as suggested by  several statistical tests they performed, a claim challenged by \citet{Moresco16}. Without entering this particular debate, we remark the significant differences between these two models at high $z$.
 In this context, we mention that \citet{Delubac15} find a 2.5 $\sigma$ difference of the BAO at $z=2.34$
 with the predictions of a flat $\Lambda$CDM model with the best-fit Planck
 parameters.
 
 Interestingly enough, the scale invariant $k=0$ and $\Omega_{\mathrm{m}}=0.30$ model is intermediate between the 
 $\Lambda$CDM and $R_h=c \, t$ models and it matches well the observations of the expansion history $H(z)$ vs. $z$ from cosmic chronometers.  In particular, we notice the good agreement with the high precision data by \citet {Delubac15}.
 
 \subsection{The values of $q_0$ in the $\Lambda$CDM and scale invariant models}

\begin{figure}[t!]
\begin{center}
\includegraphics[width=11.0cm, height=7.8cm]{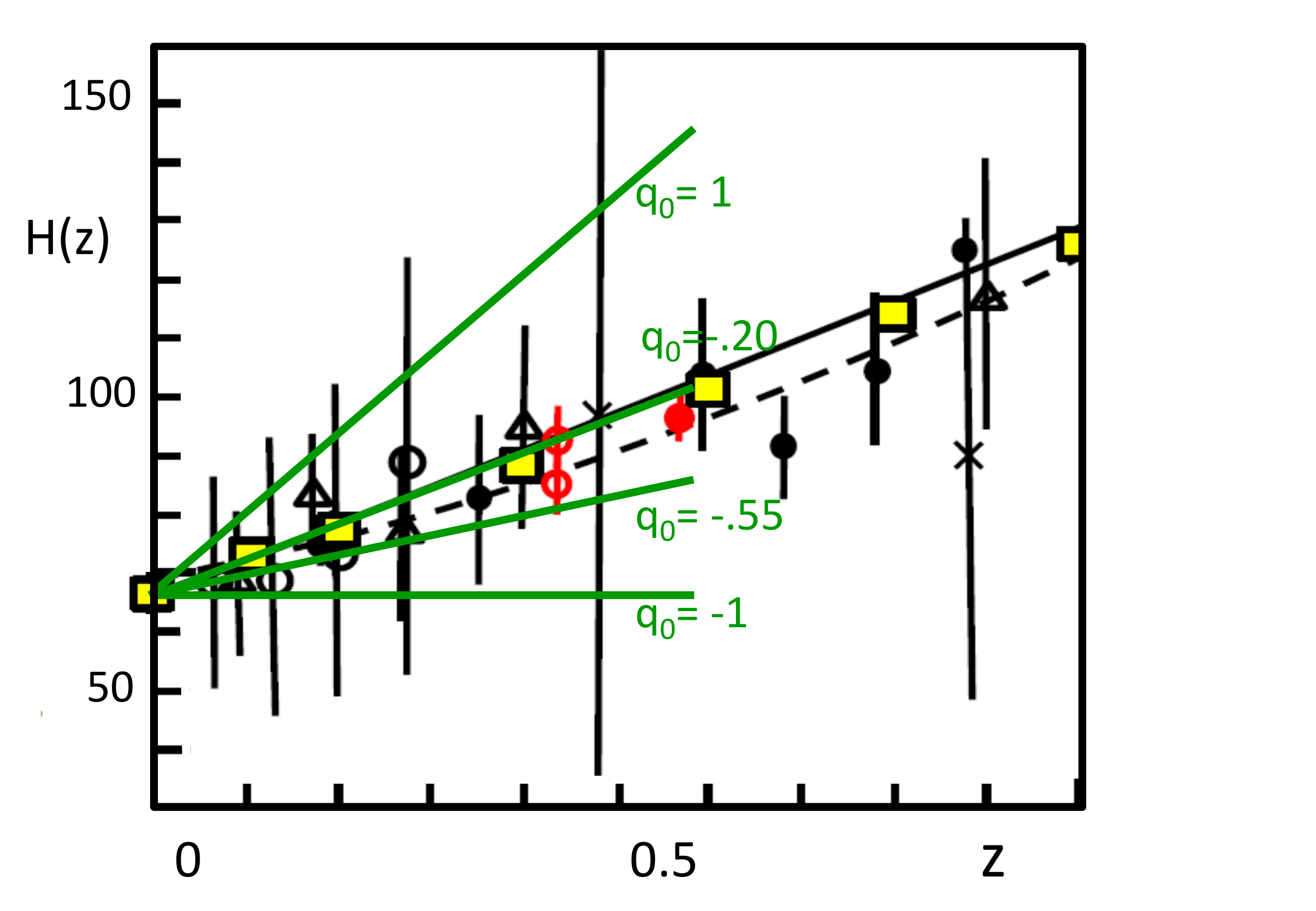}
\caption{The lower left part of   Fig. \ref{history} with the  lines indicating  the  slope $(dH/dz)_0$ for 4 different $q_0$-values.
The value $q_0=-0.55$ corresponds to the flat  $\Lambda$CDM, while  $q_0=-0.20$ for the flat scale invariant model.
For  both models, a value 
$\Omega_{\mathrm{m}}=0.30$ is assumed.}
\label{testq}
\end{center}
\end{figure}

The so-called deceleration parameter $q_0$ is testing the second derivative of $R(t)$ at $t_0$, thus it depends on the change
of the expansion rate $H$ over the recent time, {\it{i.e.}} on the values of $H(z)$ over small redshifts $z$.
As seen in Sect. \ref{flat}, the $\Lambda$CDM and the scale invariant models predict different values of the
deceleration parameter $q_0$. For $k=0$  and $\Omega_{\mathrm{m}}=0.30$, these are respectively -0.55 and -0.20, 
both corresponding to an acceleration,  slightly stronger for the $\Lambda$CDM model. The  parameter $q$ expresses 
a second derivative of $R(t)$ and is thus related to $dH/dz$, which we have 
studied in Fig. \ref{history}. We have

\begin{equation}
q=-\frac{\ddot{R} R}{\dot{R}^2}= -\frac{dH}{dt}\frac{R^2}{\dot{R}^2} -1=-\frac{dH}{dz} \frac{dz}{dt} \frac{1}{H^2} -1 \, .
\end{equation}
\noindent
In the limit $z \rightarrow 0$, we have  $-dz/dt=H_0$, thus  we get
\begin{equation}
\left(\frac{dH}{dz}\right)_0 = \, (q_0+1) H_0 \, ,
\label{qh}
\end{equation}
which relates  $q_0$ and the derivative $(dH/dz)_0$ at the present time.

Fig. \ref{testq} shows the  slopes $(dH/dz)_0$ for four different $q_0$-values, $q_0= 1, -0.20, -0.55, -1.0$. 
These slopes have to be considered in the zone near the origin $z=0$, in view of the approximations we have made.
 The differences between the various slopes are  significant.
 For a strongly decelerating Universe with $q_0=1$, we consistently see that the expansion factor $H$ 
 was much larger in the past, thus the steeper  slope in the figure. Conversely, for a moderately accelerating Universe
 the difference between  past and present values is smaller.
 We remark that both the $\Lambda$CDM and scale invariant models  for $\Omega_{\mathrm{m}}=0.30$ are within the scatter 
 of the observations, so that it would be  meaningless to speculate which one is the best. At this stage, we may  conclude that 
 the scale invariant model shows no disagreement with observations.  Maybe higher precision data may allow a
separation in the future.

\subsection{The transition from braking to acceleration}   \label{qtrans}

\begin{figure}[t!]
\begin{center}
\includegraphics[width=12.0cm, height=9.0cm]{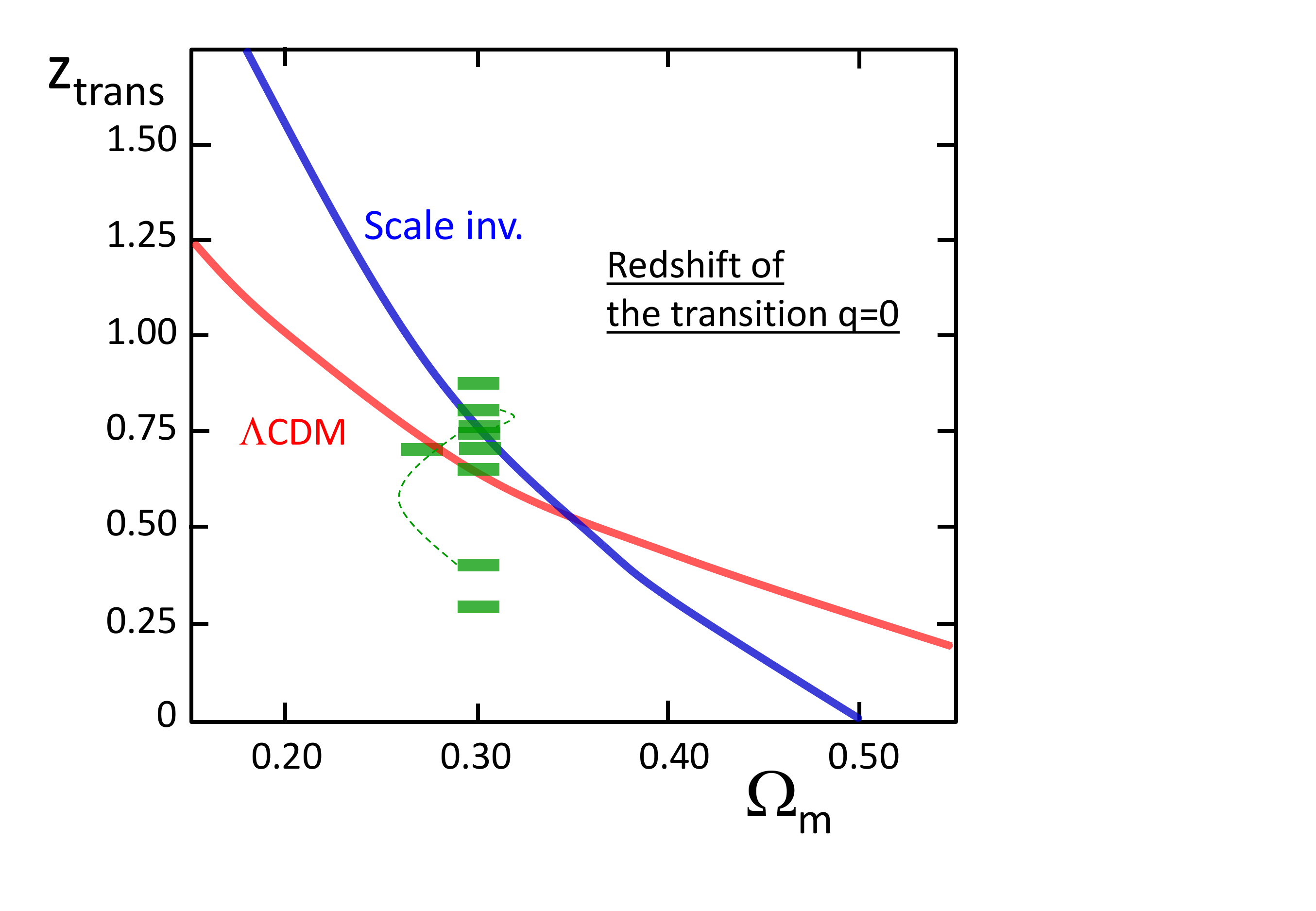}
\caption{Relation between the redshift of the transition from the braking to the acceleration of  expansion  vs. the matter density $\Omega_{\mathrm{m}}$ for the flat $\Lambda$CDM and scale invariant models.
The observational values discussed in the text  are
shown by small green rectangles.
}
\label{ztrans}
\end{center}
\end{figure}

\begin{figure*}[h!]
\centering
\includegraphics[width=.95\textwidth]{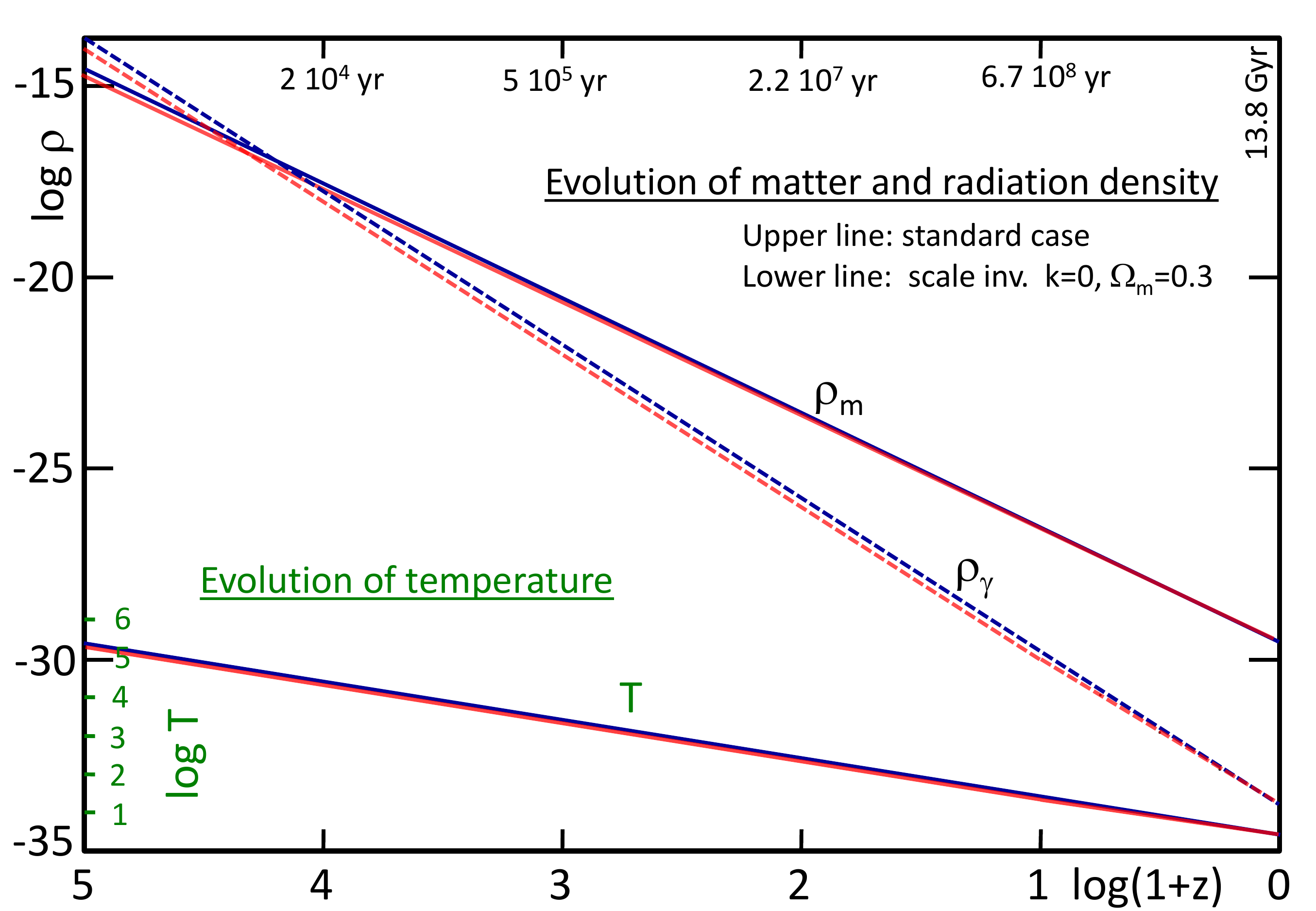}
\caption{Evolution of matter density,  radiation density and temperature as a function of redshift $z$.
 For each of the   three quantities $\rho_{\mathrm{m}}$, $\rho_{\gamma}$ and $T$, 
the upper blue line corresponds to the standard case with the classical conservation laws, while  the lower red line corresponds to the scale invariant 
solutions. For temperature, the two lines are very close to each other and cannot be distinguished in the Figure.  On the upper side of the 
frame,  the corresponding  ages  $t(z)-t_{\mathrm{in}}$ given by the flat scale invariant 
model with $\Omega_{\mathrm{m}}=0.30$  are indicated.}
\label{evolrt}
\end{figure*}

We have seen in Sect. \ref{flat} the conditions for the occurrence of the
transition from braking to acceleration which produces an inflexion point
 in the expansion $R(t)$. For the scale invariant model with $k=0$, $q=0 $ occurs
when  $\Omega_{\lambda}= \Omega_{\mathrm{m}} = \frac{1}{2}$.
 For $ \Omega_{\mathrm{m}} =0.30$, the transition occurs at $R/R_0= 0.568$ (cf. Table 1) corresponding to
a transition redshift $z_{\mathrm{trans}}=0.76$. In the $\Lambda$CDM model, the transition lies at  \citep{Suther15},
\begin{equation}
1+z_{\mathrm{trans}}=\left( \frac{2\, \Omega_{\Lambda}}{\Omega_{\mathrm{m}} }\right)^{1/3} \, ,
\end{equation}
\noindent
so that for the same $\Omega_{\mathrm{m}} $, one has $z_{\mathrm{trans}}=0.67$, i.e. slightly later in the expansion.
 Fig. \ref{ztrans} shows as a function of    $\Omega_{\mathrm{m}} $ 
the values of the redshift $z_{\mathrm{trans}}$ at which the transitions are  located for both the $\Lambda$CDM and the scale invariant models.
   $z_{\mathrm{trans}}$  varies faster with matter density for the scale invariant 
 than for  the $\Lambda$CDM case. However,  the two curves are crossing at about a matter density
 $\Omega_{\mathrm{m}} \approx 0.35$ so that they are still rather close to each other at $\Omega_{\mathrm{m}} = 0.30$.
 The distinction of the two cases may be possible in the future with accurate data, for now it is still uncertain.

Since a decade, several authors have tried to estimate the value of $z_{\mathrm{trans}}$. This is a difficult task, since
it concerns the second derivative of $R(t)$, implying the study of the change of $H(z)$ with redshift $z$. In addition, the estimates
are often not model independent and this may introduce a bias in the comparisons. 
The study by 
\citet{Shap06} suggested that   the transition  $z_{\mathrm{trans}}$ lies at $ \approx 0.3$ for $\Omega_{\mathrm{m}}=0.30$,
 a value of the
matter-density adopted in most studies below.
\citet{Melch07} found a much higher value, then generally also supported by the followers. Depending on different assumptions
concerning the equation of state,  these authors obtained a value of $z_{\mathrm{trans}}$ between $0.76 \pm 0.10$ and 0.81$\pm 0.12$,
implying that the transition occurred 6.7 Gyr ago (resp. 6.9 Gyr). The two values are connected by a thin broken line in 
Fig. \ref{ztrans}. 
 \citet{Ishida08} from  data on supernovae, on the CMB and BAO,
found a value $z_{\mathrm{trans}}=0.88 \; (+.12, -.10)$. 
\citet{Blake12} gave $z_{\mathrm{trans}} \approx 0.7$ for $\Omega_{\mathrm{m}}=0.27$ 
A recent analysis  by \citet{Suther15} indicates that the SN data are better for the estimate of the acceleration over recent epochs,
while BAO measurements   may more constrain the value of $z_{\mathrm{trans}}$. They suggest  $z_{\mathrm{trans}} \sim 0.7 $.
\citet{Rani15} apply a model independent approach with different parameterizations, which all support a value $z_{\mathrm{trans}} < 1.0$,
 with a likely value around 0.7. \citet{Viten15} generate by Monte-Carlo methods mock catalogs and compare them to observations
to determine the transition. Their best fit supports a transition redshift  $z_{\mathrm{trans}} \approx 0.65$. \citet{Moresco16} find a value 
 $z_{\mathrm{trans}}=0.4 \pm 0.1$ for one of the models of spectral evolution they use, while for another model they get
  $z_{\mathrm{trans}}=0.75 \pm 0.15$. The two results are connected by a thin broken line in 
Fig. \ref{ztrans}.

We see that most of the estimates support a transition near
$z_{\mathrm{trans}}= 0.75$, except two. One was the first work on the topic \citep{Shap06}, the other by \citet{Moresco16}
depends on the adopted model for spectral evolution chosen. On the whole, the observations are in  good agreement with the flat scale invariant
models with $k=0$.  However, the differences at  $\Omega_{\mathrm{m}}=0.30$ 
between the  $\Lambda$CDM and the scale invariant model in Fig. \ref{ztrans} are  small and not sufficient to discriminate between
the two models.
  
Moreover, we note that the transition in the models from braking to acceleration is not a sharp 
and strong one (e.g. Fig. \ref{Rtzero}), the two phases being separated 
by a non negligible transition phase where $R(t)$ is almost linear. This contributes to make the observational determination
of  $z_{\mathrm{trans}}$ a difficult challenge.

\section{Past evolution  of matter density, radiation density and temperature}

We want to start  examining the past evolution of the matter and radiation densities, as well as of the  temperature $T$ in the scale invariant model to see what may be the changes in the past history of the Universe predicted by the scale invariant models. We may wonder 
about the changes,
especially more than the conservation law (\ref{3w}) contains a $\lambda$-term which  leads to differences with respect to the standard case. According to (\ref{3w}), the matter and radiation densities $\varrho_{\mathrm{m}}$ and $\varrho_{\gamma}$ with respectively  $w=0$ and $w=1/3$  obey the relations,
 
\begin{equation}
\varrho_{\mathrm{m}} \, R^3 \, \lambda  = \mathrm{const} \quad \quad \mathrm{and} \quad 
\varrho_{\gamma} \, R^4 \, \lambda^2 = \mathrm{const'}  \, .
\label{roro}
\end{equation}
\noindent
Since $\varrho_{\gamma}$ behaves like $T^4$,  the temperature of cosmic microwave background is determined by

\begin{equation}
T \, R \, \lambda^{\frac{1}{2}} \, = \, \mathrm{const''} \, .
\label{T}
\end{equation}
\noindent
Fig. \ref{evolrt} shows the past evolution of these quantities versus redshift with the  scale $\log(1+z)$. For the present value,  we take 
$\log \varrho_{\mathrm{m}}=-29.585$ corresponding to $\Omega_{\mathrm{m}}=0.30$ and $H_0=67.8 $ km s$^{-1}$ Mpc$^{-1}$,
as given by the Planck Collaboration (see Sect. \ref{Hh}). For the present temperature, we take $T_0=2.726$ \citep{Fix09}.
This leads to a radiation density  $\log \varrho_{\gamma }=-33.768$. The values of the $\lambda$-parameter are obtained from Table
\ref{detailk0} in the Appendix.  A few values of the cosmic time are given on the upper line of the frame for the reference model. The above expressions (\ref{roro}) and ({\ref{T}) show that as $\lambda$ was bigger in the past
(unlike $R(t)$), the values of  $\varrho_{\mathrm{m}}$, $\varrho_{\gamma}$ and $T$ for the scale invariant cosmology
were lower than those given by the standard case. 

Amazingly, the differences between the scale invariant and the standard case are very small. The reasons are the following ones.
As illustrated by Fig. 1, $R(t)$ decreases very  rapidly  (thus making a large increase of redshift $z$) for a small change of  $t/t_0$ (and thus of $\lambda$, see  Table \ref{detailk0}).  Also, we have seen in Fig. \ref{scale} that the domain of 
the variations of the $\lambda$-parameter is  limited  to values between 1.0 (now) and about 1.5 at 
the Big-Bang for     $\Omega_{\mathrm{m}}=0.30$. 
For a density parameter $\Omega_{\mathrm{m}}$  closer to 1.0, the differences between the curves in 
Fig. \ref{evolrt} would even be smaller.
  On the whole, the evolution 
of matter and radiation densities is very similar, although not strictly identical,  to the result of  the standard case given by the classical conservation laws. A  calendar 
giving  times $t$ as a function of redshift is given by Table \ref{detailk0} for the reference scale invariant model.

The crossing of the two curves $\varrho_{\mathrm{m}}$ and $\varrho_{\gamma}$ indicating the transition from the matter dominated era to the radiation era occurs at

\begin{equation}
\log (1+z)_{\mathrm{cross}}=4.183, \,\frac{R}{R_0}_{\mathrm{cross}}= 6.5615 \;10^{-5},
\frac{t}{t_0}_{\mathrm{cross}}= .6694288\, .
\label{cross}
\end{equation}
\noindent
The difference in the  redshifts of the crossing   for the standard case of evolution and the scale invariant model
with  $\Omega_{\mathrm{m}}=0.30$ is very small as illustrated by Fig. \ref{evolrt}.
As the origin $R(t_{\mathrm{in}})=0$ lies at ${t_\mathrm{in}}_/t_0= 0.6694285$, the age of the crossing is about $ 4 \cdot 10^3$ yr.
 
During the radiation era, the dominant equation of state is different from that in the present matter era, 
 thus the cosmological equations and their solutions are different. The exploration of the radiation era is beyond the scope of the present work, especially more
than at some very early stage, the assumption of scale invariance of the empty space should break down. Nevertheless,
 we may wonder whether the origin $t_{\mathrm{in}}$ of the Universe, predicted for this era, occurs at about  the time that
  we have derived in Sect. \ref{flat}. If this not the case, we would have to change the origin $t_{\mathrm{in}}$ that we have used above. To check this point, we must integrate equation (\ref{E1}) with the appropriate conservation law.
 Equation (\ref{E1}) becomes
for $k=0$ and  with  $\varrho_{\gamma} \, R^4 \, \lambda^2= \mathrm{const}$, 

\begin{equation}
\frac{8 \, \pi G \varrho \, R^4 \lambda^2 }{3} = \dot{R}^2 R^2 \lambda^2 +2 \,\dot{R} R^3 \dot{\lambda} \lambda \,  .
\end{equation} 
\noindent
Calling $C_{\mathrm{rad}}$ the first member of the above equation, we get expressing $\lambda$ with (\ref{lamb})

\begin{equation}
\dot{R}^2 R^2 \, t - 2 \, \dot{R}\, R^3  - C_{\mathrm{rad}} \, t^3 =0 \, ,
\label{Ex}
\end{equation}
which can be compared to the equation  (\ref{E13}) of the matter dominated era. We have to express the constant $C_{\mathrm{rad}}$. At the crossing point, 
we have identical values 
of $R(t)$, $\lambda(t)$ and by definition we also have the equality $\varrho_{\mathrm{m}}=\varrho_{\gamma}$, this implies

\begin{equation}
C_{\mathrm{rad}} \, = \, C \ \ R_{\mathrm{cross}} \, \lambda_{\mathrm{cross}} \, , 
\end{equation}

\noindent
where $C$ is the value used in (\ref{E13}).
Numerically, with the value of $C=2.44898$ for the reference model, we get $C_{\mathrm{rad}}= 2.40040 \cdot 10^{-4}$. We may thus proceed to the integration during the radiation era. 
We check here that the origin we may determine from (\ref{Ex}) brings no significant change in the age scale we have adopted above.
 The integration of (\ref{Ex}) leads to  a value of $t_{\mathrm{in}}/t_0$ that differs  by less than the last 
  digit of that obtained for the crossing time.
 Thus, for the present purpose, we may keep the same origin at that found previously, see Tables 1 and 3.

There is, however an interesting difference. Relation (\ref{Ex}) imposes an extremely fast initial expansion during the  radiation era. 
The initial  rate  $\dot{R}$ tends towards infinity at the origin,
  this even more applies to the Hubble term $H$  near the origin. This  suggests    that the scale invariant models containing matter
  experience a Big-Bang.   However, at the level of quantum physics in   the most early stages, the assumption of the scale invariance of the empty space   likely breaks down and a more appropriate physics  would be needed to treat  this event.

\section{Conclusions}

There are strong physical motivations to enlarge the group of invariances 
sub-tending the theory of gravitation and cosmology. In this context, the specific  hypothesis we have made 
 about the  scale invariance of the empty space at large scales  seems to open a window on possible interesting 
 new cosmological models.  The various comparisons of models and observations we have made so far
 on the dynamical properties of the scale invariant cosmology  are positive and thus 
encouraging  for the continuation of the investigations. 
If true, the hypotheses we made have many other implications in astrophysics. Thus,
 these cosmological models evidently  need 
 to be further thoroughly  checked with many other possible astrophysical tests.

In view of further tests, a point  about methodology needs to be strongly  emphasized:  
to be valid, a test must be internally coherent and make no use of properties or inferences  from the framework of other cosmological  models, a point which is not always evident.

\vspace*{3mm}

\noindent
Acknowledgments: I want to express my best thanks to the physicist D. Gachet and Prof. G. Meynet for their 
continuous encouragements.

\appendix

\section{Details of the scale invariant model with $k=0$ and $\Omega_{\mathrm{m}}=0.30$}

\begin{table*}[h!]  
\label{detailk0}
\vspace*{0mm}
 \caption{Data of the reference scale invariant model with $k=0$  and $\Omega_{\mathrm{m}}=0.30$.}
\begin{center}  \small
\begin{tabular}{cccccccc}
$z$  & $R/R_0$  & $t/t_0$  & $ \tau/t_0  $ & age   &  $H(t_0)$ &    $H(z)$   &   $\lambda$ \\
      &              &             &                   &(yr)    &               & km s$^{-1}$ Mpc$^{-1}$& \\
\hline
 &   &   &   \\
0.00    &   1       &      1     &  .3306     & 13.8 E+09 &  2.857  &   67.0 & 1.000    \\
0.05    & .9524   &  .9833   &  .3139     & 13.1 E+09 &  2.972   &   69.7 & 1.017    \\
0.10    & .9091   &  .9679   &  .2985     & 12.5 E+09 &  3.088   &   72.4  & 1.033   \\
0.20    & .8333   &  .9407   &  .2713     & 11.3 E+09 &  3.324   &   77.9  & 1.063   \\
0.40    & .7143   &  .8974   &  .2280     & 9.5 E+09  &  3.810   &   89.4   & 1.114 \\
0.60    & .6250   &  .8644   &  .1950     & 8.1 E+09 &   4.321   &  101.3  &  1.157  \\
0.80    & .5556   &  .8387   &  .1693     & 7.1 E+09 &   4.852   &   113.8 &  1.192   \\
1.00    & .5000   &  .8181   &  .1487     & 6.2 E+09 &   5.408   &   126.8  & 1.222   \\
1.20    & .4545   &  .8013   &  .1319     & 5.5 E+09 &   5.987   &   140.4  & 1.248   \\
1.50    & .4000   &  .7814   &  .1120     & 4.7 E+09 &   6.895   &   161.7  &  1.280 \\
2.00    & .3333   &  .7575   &  .0881     & 3.7 E+09 &   8.522   &   199.9  &  1.320   \\
3.00    & .2500   &  .7290   &  .0596     & 2.5 E+09 &   12.16   &   285.1  &  1.372   \\
4.00    & .2000   &  .7131   &  .0437     & 1.8 E+09 &   16.24   &   381     &  1.402\\
6.00    & .1429   &  .69642 &  .0270     & 1.1 E+09 &    25.67   &   602    &   1.4359   \\
9.00    & .1000   &  .68550 &  .0161     & 6.7 E+08 &    42.46   &   996    &   1. 4588\\
99      & .0100   &  .66995 & 5.3 E-04   & 2.2 E+07 &  1.28 E+03   &    3.0 E+04 & 1.4926    \\
999    & .0010   &  .66944  &1.2  E-05  & 4.8 E+05 &   4.08 E+04 &    9.6 E+05   & 1.4938  \\
9999   & .0001  &  .66943  & 5.0  E-07  & 2.1 E+04 &  1.27 E+06 &    3.0 E+07    & 1.4938 \\
\hline
\normalsize
\end{tabular}
\end{center}
\end{table*}

In Table \ref{detailk0}, we give same basic data for the reference model  with $k=0$ and $\Omega_{\mathrm{m}}=0.30$
as a function of the redshift $z$. 
Column 2 gives the solution $R(t)$ of Eq. (\ref{E13}) 
for different values of the time $t/t_0$ (column 3). Column 4 contains the age $\tau= t-t_{\mathrm{in}}$.  
The present age in year is given in column 5 for a present value of 13.8 Gyr.
Column 6 gives the Hubble parameter $H(t_0)$ in the scale $t_0=1$, while the Hubble parameter $H(z)$ in km s$^{-1}$ Mpc$^{-1}$ is given in 
column 7 for  the same assumption about the age of the Universe of 13.8 Gyr  as in Table 1. In column 8, the scale factor
$\lambda$ is given with $\lambda=1$ at present.

\bibliographystyle{aa}
\bibliography{Maeder III-AA}

\end{document}